\documentclass[preprint]{aastex631}

\shorttitle{Spectral Characteristics of FRB 20201124A}
\shortauthors{Lyu, Liang, and Li}
\graphicspath{{./}{figures/}}

\usepackage{footnote}
\usepackage{amsmath}
\usepackage{threeparttable}
\usepackage{diagbox}
\begin{document}
\title{Narrowly-Banded Spectra with Peak Frequency Around 1 GHz of FRB 20201124A: Implications for Energy Function and Radiation Physics
}

\author[0000-0002-6072-3329]{Fen Lyu}
\affiliation{Institute of Astronomy and Astrophysics, Anqing Normal University, Anqing 246133, People’s Republic of China}
\affiliation{Guangxi Key Laboratory for Relativistic Astrophysics, School of Physical Science and Technology, Guangxi University, Nanning 530004, China}
\author[0000-0002-7044-733X]{En-Wei Liang*}\thanks{E-mail: lew@gxu.edu.cn}
\affiliation{Guangxi Key Laboratory for Relativistic Astrophysics, School of Physical Science and Technology, Guangxi University, Nanning 530004, China}
\author[0000-0003-3010-7661]{D. Li*}\thanks{E-mail: dili@nao.cas.cn}
\affiliation{CAS Key Laboratory of Radio Astronomy, National Astronomical Observatories, Chinese Academy of Sciences, Beijing, China}
\begin{abstract}
The radiation physics of fast radio bursts (FRBs) remains an open question. Current observations have discovered that narrowly-banded bursts of FRB 20201124A are active in 0.4-2 GHz and their spectral peak frequency ($\nu^{\rm obs}_{p}$) are mostly toward $\sim 1$ GHz. Utilizing a sample of 1268 bursts of FRB 20201124A detected with the FAST telescope, we show that the $1\sigma$ spectral regime of 71.4\% events (in-band bursts) is within the FAST bandpass. Their intrinsic burst energies ($E^{\rm obs}_{\rm BWe}$) and spectral widths ($\sigma_s^{\rm obs}$) are well measured by fitting the spectral profile with a Gaussian function. The derived $E^{\rm obs}_{\rm BWe}$ and $\sigma_s^{\rm obs}$ distributions are log-normal and centering at $\log E^{\rm obs}_{\rm BWe}/{\rm erg}=37.2~ (\sigma=0.76)$ and $\log \sigma_s^{\rm obs}/{\rm GHz}=-1.16~  (\sigma=0.17)$. Our Monte Carlo simulation analysis infers its intrinsic $\nu_p$ distribution as a normal function centered at $\nu_{p,c}=1.16$ GHz ($\sigma=0.22$) and its intrinsic energy function as $\Phi(E)\propto E^{-0.60}e^{-E/E_c}$ with $E_c=9.49 \times 10^{37}$ erg. We compare these results with that of typical repeating FRBs 20121102A and 20190520B that are active over a broad frequency range at several specific frequencies and discuss possible observational biases on the estimation of the event rate and energy function.  Based on these results, we argue that FRB 20201124A likely occurs in a fine-tuned plasma for maser radiations at a narrow frequency range, while FRB 20121102A and FRB 20190520B could involve clumpy plasma conditions that make maser emission around several specific frequencies in a broad range.
\end{abstract}

\section{Introduction} \label{sec:intro}
As extremely short (the typical duration of $\mu s-$ms) and bright (the typical fluence of $\sim$Jy ms) radio flashes discovered in 2007 \citep{2007Sci...318..777L}, fast radio burst (FRB) phenomenon is still a great puzzle, despite significant progress has been made in recent years with benefits from both radio surveys for accumulating the FRB sample and extensive observations for individual repeating FRBs (see \citealt{2019ARA&A..57..417C,2019A&ARv..27....4P,2020Natur.587...45Z} for reviews). More than 800 FRBs have been reported in frequency from hundreds of MHz to several GHz\footnote{https://www.herta-experiment.org/frbstats/catalogue}. Most FRBs are of extragalactic origin. This was firstly inferred from their large dispersion measure (DM) excesses over the maximum distribution of the free electrons estimated in the NE2001 and YMW16 Galactic disk models \citep{2002astro.ph..7156C,2017ApJ...835...29Y}, and now has been confirmed by the host galaxy localization for several FRBs (e.g. \citealt{2017Natur.541...58C,2019Sci...365..565B,2019Natur.572..352R}). The great majority of FRBs are non-repeaters (also called one-off FRBs) even after long-term monitoring, while a few fractions of FRB sources exhibit repeating behaviors \citep{2021ApJS..257...59C,2023ApJ...947...83C}. It is uncertain whether the repeating FRBs are different from the one-off FRBs \citep{2021ApJ...906L...5A,2022ApJ...939...27C}. Statistical analysis for a sample of 536 FRBs (including 18 repeating FRBs and 474 apparent non-repeaters) detected with the CHIME telescope demonstrates that the repeating FRBs tend to have wider temporal width and narrower spectral bandwidth than those of one-off FRBs \citep{2021ApJS..257...59C}. The bursts of the repeating FRBs typically exhibit sub-bursts and frequency drifts \citep{2021ApJ...923....1P}. For seeking the burst nature, analysis of the energy function and event rate for a given repeating FRB source and/or from a sample of one-off FRBs has been extensively investigated (e.g. \citealt{2019JHEAp..23....1D,2019ApJ...883...40L,2019ApJ...882..108W,2019MNRAS.487.3672Z,2020ApJ...895L...1D,2020MNRAS.494..665L,2021Natur.598..267L,2021FrPhy..1624503L,2021ApJ...920L..23Z,2021MNRAS.501..157Z,2021ApJ...920..153W,2022MNRAS.511.1961H,2022RAA....22l4002Z,2023arXiv231000908Z}). However, the origin and radiation physics of FRBs are still under debate.

Repeating FRBs offer the opportunity to gain insight into the intrinsic nature of the FRB phenomenon through extensive observations across different observing frequencies. FRB 20121102A is the first one among them and has been extensively monitored with multiple telescopes at the frequencies from 0.5 to 8 GHz (e.g. \citealt{2014ApJ...790..101S,2016ApJ...833..177S,2016Natur.531..202S,2017Natur.541...58C,2017ApJ...834L...7T,2017ApJ...846...80S,2018Natur.553..182M,2018ApJ...866..149Z,2019ApJ...877L..19G,2019A&A...623A..42H,2019ApJ...882L..18J,2020MNRAS.496.4565C,2020ApJ...891L...6F,2020ApJ...897L...4M,2020MNRAS.495.3551R,2020A&A...635A..61O,2021Natur.598..267L,2022MNRAS.515.3577H,2021ApJ...922..115A,2023MNRAS.522.3036C,2023MNRAS.519..666J}). It is interesting that a tentative bimodal distribution of the isotropic equivalent specific energy ($E_{\mu_{\rm c}}$) is found by \cite{2021Natur.598..267L} from 1652 bursts observed with the Five-hundred-meter Aperture Spherical radio Telescope (FAST, \citealt{2018IMMag..19..112L,2020RAA....20...64J}), where $\mu_{\rm c}$ is the central frequency of the instrument. The emission spectra of the bursts from FRB 20121102A are narrow, and their profile can be described with a Gaussian function \citep{2017ApJ...850...76L,2018ApJ...863....2G,2019ApJ...882L..18J}. Analyzing the data observed with the Green Bank Telescope (GBT) in the $C$-band (4-8 GHz) for FRB 20121102A, \cite{2022ApJ...941..127L} found that the distribution of the burst emission peak frequencies ($\nu_p$) illustrates some discrete peaks. Extrapolating this $\nu_p$ distribution feature to 0.5-4 GHz, they demonstrated that the bimodal $E_{\mu c}$-distribution of FRB 20121102A in the FAST sample can be reproduced via Monte Carlo simulations by assuming a single power-law energy function. The discrete $\nu_p$ distribution is also favored by the fact that the bursts of FRB 20121102A seem to be active in some specific frequencies (e.g. \citealt{2016Natur.531..202S,2021Natur.598..267L,2022MNRAS.515.3577H}). FRB 20190520B is another active repeating FRB. It shares many similar features with FRB 20121102A. Utilizing the data of FRB 20190520B observed with the GBT, VLA, and Parkes telescopes  \citep{2022arXiv220308151D,2022Sci...375.1266F,2022Natur.606..873N,2023Sci...380..599A}, \cite{2023MNRAS.522.5600L} identified a similar discrete $\nu_p$ distribution over the frequency of $\sim$1-6 GHz.

FRB 20201124A is another very bright and exceptionally active repeating FRB source. It was initially detected by \cite{2021ATel14497....1C} in the frequency of 400-800 MHz on March 31, 2021, and subsequently located by the Australian Square Kilometre Array Pathfinder (ASKAP, \citealt{2021ATel14515....1D}), $\text {VLA/realfast}$ \citep{2021ATel14526....1L}. Its DM value is $413.5 \pm 0.5$ pc cm$^{-3}$, and its host galaxy was identified as a metal-rich, barred-spiral galaxy SDSS J050803.48+260338.0 at z=0.098 \citep{2020ApJ...899..161L,2021ApJ...919L..23F,2022MNRAS.513..982R,2022Natur.609..685X}. An associated extended persistent radio source (PRS) is also discovered by the upgraded Giant Metrewave Radio Telescope (uGMRT) at 550-750 MHz, VLA at 3 GHz and 9 GHz \citep{2021ATel14549....1R,2021ATel14529....1W,2022MNRAS.513..982R}. Conducting a simultaneous multi-wavelength survey in the radio, optical, and X-ray bands, \cite{2021A&A...656L..15P} found that this FRB source located at the center of the PRS associated with the brightest region of a star-forming region in the host galaxy. With observations of the Effelsberg 100-m radio telescope, the ultra-wideband low (UWL) receiver at the Parkes 64-m radio telescope, and the FAST, \cite{2021MNRAS.508.5354H}, \cite{2022MNRAS.512.3400K} and \cite{2022Natur.609..685X} found that the bursts of FRB 20201124A exhibit an irregular and short-time variation of the Faraday rotation measure and high polarization.

Thousands of the bursts of FRB 20201124A have been detected with radio telescopes in the frequency range of 0.4-2.0 GHz (e.g. \citealt{2022RAA....22j5007L,2022ApJ...927L...3N,2022MNRAS.513..982R,2022RAA....22l4001Z,2023PASJ...75..199I}). Most of them are observed with the FAST in $1\sim 1.5$ GHz. \citep{2022Natur.609..685X,2022RAA....22l4001Z,2022RAA....22l4002Z,2022RAA....22l4003J,2022RAA....22l4004N}. In addition, 5 bursts were observed by the Parkes UWL receiver over the range of 704-4032 MHz at $\leq$ 1.1 GHz \citep{2022MNRAS.512.3400K}, 20 bursts were observed by the 100m Effelsberg radio telescope at 1.27-1.45 GHz \citep{2021MNRAS.508.5354H}, 18 bursts were detected by VLBI at 1254-1510 MHz over two epochs \citep{2022ApJ...927L...3N}, 3 bursts were detected with the Xinjiang Nanshan 26-m radio telescope at 1396-1716 MHz
\citep{2022ATel15289....1Y}, and 15 bursts observed in the ASKAP mid-band at 1103.5-1439.5 MHz \citep{2021ApJ...919L..23F,2022MNRAS.512.3400K}. At the low frequency below 1 GHz, about one burst was detected in the ASKAP low-band (696.5-1032.5 MHz) per every observational session \citep{2021ApJ...919L..23F,2021ATel14508....1K,2021ATel14502....1K}. Thirty-three bursts were detected with the CHIME telescope at 400-800 MHz \citep{2022ApJ...927...59L}, and 48 bursts were detected with the uGMRT telescope at 550-750 MHz \citep{2022MNRAS.509.2209M}. No burst has been detected at the frequency below $0.4$ GHz, despite efforts for observation with the Murchison Widefield Array (MWA) in 144-215 MHz \citep{2023MNRAS.518.4278T}. Only one burst was observed at 2 GHz, but no burst was detected in the $X$-band (8374-8502 MHz) by simultaneously monitoring with the 64-m radio dish of Usuda Deep Center$/$JAXA \citep{2023PASJ...75..199I}. Furthermore, no burst was detected at the frequency above 2 GHz with the Deep Space Network \citep{2021ATel14519....1P}. We should note that the detection heavily depends on the sensitivity of the instrument and the amount of time spent in follow-up monitoring in each frequency band. The current observations indicate that FRB 20201124A is only active at a frequency below 2 GHz and mostly toward the lower end of the L-band. This is dramatically different from that of FRB 20121102A and FRB 20190520B.

Similar to other typical repeating FRBs, the bursts emitted from FRB 20201124A also exhibited narrow-banded spectra \citep{2022MNRAS.512.3400K,2022RAA....22l4001Z} and the frequency downward drift \citep{2021MNRAS.508.5354H,2022RAA....22l4001Z}. It is well known that FRBs could be from coherent radiation based on their high brightness temperature (see \citealt{2019A&ARv..27....4P,2020Natur.587...45Z,2021SCPMA..6449501X} for reviews). The currently proposed models can be grouped into two general classes: coherent curvature radiation in bunches (bunch emission; e.g., \citealt{2016ApJ...829...27D,2018ApJ...852..140W,2018ApJ...868...31Y,2020MNRAS.498.1397L,2022SCPMA..6589511W}) and maser emission (negative absorption; e.g., \citealt{2014MNRAS.442L...9L,2017ApJ...843L..26B,2019MNRAS.485.4091M,2020ApJ...896..142B,2020ApJ...899L..27M}). In the bunch emission scenario, the bursts are thought to be coherent curvature radiations by bunches of charged particles inside the magnetosphere. It predicts wide emission spectra ($\Delta \nu /\nu \sim 1$). This cannot explain the observed narrow spectra, although the absorption of low-frequency radio emission may lead to relatively narrow spectra \citep{2018ApJ...868...31Y}. In addition, the bunch mechanism finds it hard to explain the formation and maintenance of these bunches (e.g. \citealt{2019ApJ...885L..24C}). In the maser emission mechanism scenario, the maser may emit a narrow radio spectrum in either vacuum or plasma in fine-tuned conditions. The vacuum synchrotron maser requires that the field line have a high degree of order in the emitting region, and the emitting electrons must have the same narrow distribution of pitch angle and energies \citep{2017MNRAS.465L..30G}. The plasma synchrotron maser involves a weakly magnetized plasma ($\omega_{p}>\omega_{B}$; \citealt{2017ApJ...842...34W}) or highly magnetized shocks (e.g. \citealt{2014MNRAS.442L...9L,2017ApJ...843L..26B,2019MNRAS.485.4091M,2019MNRAS.485.3816P,2020ApJ...896..142B,2020ApJ...897....1L}). The high event rate and extremely narrow frequency coverage of the bursts in the L-band from FRB 20201124A provide a great advantage for exploring the entire spectral profiles of a large and uniform burst sample observed with the FAST telescope. Moreover, the bursts of FRB 20201124A are highly polarized. For more than $90 \%$ bursts in the FAST sample reported in \cite{2022RAA....22l4003J}, the total degree of polarisation (linear plus circular) is greater than $90 \%$. The high circular polarization degree might be attributed to the off-beam observation \citep{2022MNRAS.517.5080W,2023ApJ...943...47L,2023MNRAS.522.2448Q}. The emitted narrow spectrum, together with the very high total degree of polarisation \citep{2022Natur.609..685X,2022RAA....22l4003J} suggests that the observed spectrum represents the intrinsic properties of this FRB source. This can shed light on the radiation physics of these bursts.

This paper investigates the intrinsic spectral properties and energy function of FRB 20201124A via Monte Carlo simulations by using a large and uniform burst sample observed with the FAST telescope from \cite{2022RAA....22l4001Z}. It is structured as follows. The sample selection and data for our analysis are presented in Section \ref{sec:Sample}. We study the cumulative distribution of the intrinsic energy function (E-Function) and radiation properties in Section \ref{sec:sim}. Discussion and summary are present in Sections \ref{sec:Discuss} and \ref{sec:Summary}, respectively. Throughout, we adopt a flat $\Lambda$CDM universe with the cosmological parameters $H_{0}$=67.7$\mathrm{~km}$ $\mathrm{~s}^{-1}$ $\mathrm{Mpc}^{-1}$, $\Omega_{m}=0.31$ \citep{2016A/&A...594A..13P}.

\section{Sample Selection and Data Analysis}
\label{sec:Sample}
\begin{figure}[!htbp]
\centering
\includegraphics[width=0.65\linewidth]{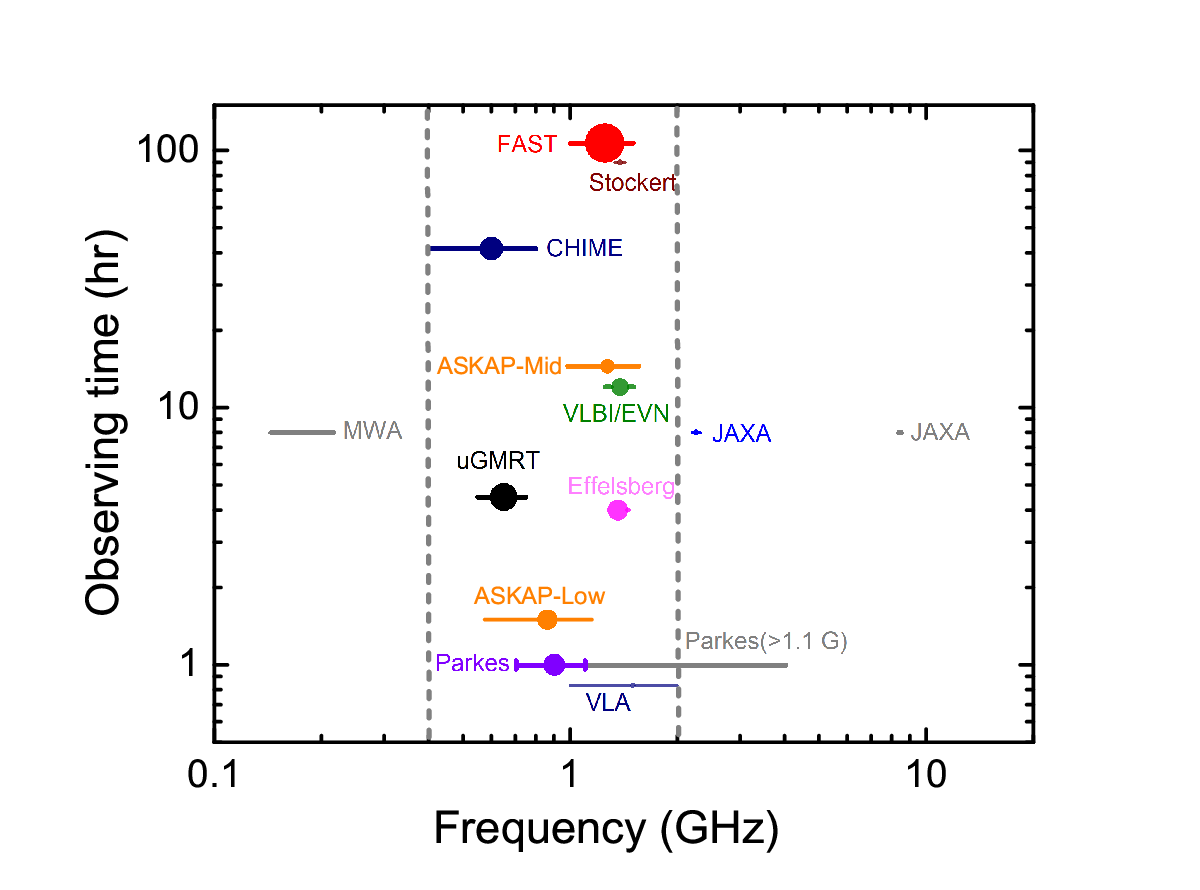}\vspace{-0.1in}
\caption{The total follow-up observation time for FRB 20201124A against the observing frequency with different telescopes in different frequency bands as marked in the plot: CHIME \citep{2022ApJ...927...59L}, ASKAP \citep{2022MNRAS.512.3400K}, FAST \citep{2022RAA....22l4004N,2022RAA....22l4003J,2022Natur.609..685X,2022RAA....22l4002Z,2022RAA....22l4001Z}, uGMRT \citep{2022MNRAS.509.2209M}, VLA \citep{2021ATel14526....1L}, 100m Effelsberg \citep{2021MNRAS.508.5354H}, UDSC/JAXA \citep{2023PASJ...75..199I}, MWA \citep{2023MNRAS.518.4278T}, VLBI/EVN \citep{2022ApJ...927L...3N}, Parkes UWL receiver \citep{2022MNRAS.512.3400K}, and the Stockert 25 m Radio Telescope \citep{2021ATel14556....1H}. The short dashed lines mark the frequency range of the bursts emitted (0.4-2 GHz, except for one extremely bright burst detected with the JAXA telescope at 2.1-2.3 GHz with peak flux $> 189$ Jy ms). The size of the circle illustrates the relative detected burst numbers, and the horizontal segments mark the current observed frequency range for FRB 20201124A.
\label{fig:various}}
\vspace{-0.2cm}
\end{figure}

As mentioned in Section 1, the bursts emitted from FRB 20201124A have been reported in multiple observations with various telescopes so far. Figure \ref{fig:various} displays the total follow-up observation time against the observing frequency in the various targeted monitoring campaigns for FRB 20201124A. The observed bursts emitted from this source are distributed in the observing frequency range of 0.4-2.0 GHz, except for one extremely bright burst observed in the S-band (2.1-2.3 GHz) with the 64-m radio dish of Usuda Deep Center/JAXA \citep{2023PASJ...75..199I}. The peak flux of this event is $>189$ Jy ms at 2 GHz, being brighter than the brightest one among the 2500 events in the FAST sample by one order of magnitude. It is quite similar to a one-off event but does not look like a normal outburst from FRB 20201124A. Except for this event, during the observing session of 8 hour duration at frequencies above 2 GHz, no bursts have been detected \citep{2023PASJ...75..199I}. In addition, no bursts have been discovered with the observations for FRB 20201124A in the low-frequency ($\nu_{p}<$0.4 GHz). It seems that FRB 20201124A is active in the frequency range of 0.4-2 GHz. However, current observations still cannot exclude the possibility that this is due to a selection bias dependent on the amount of time dedicated to searching for bursts at different frequencies.

The overwhelming majority ($>2000$ bursts) of the detected bursts reported for FRB 20201124A were monitored with the FAST telescope at the $L$-band (1.0-1.5 GHz) during the two extremely active episodes in 2021. In the first episode from April 1 to June 11, 1,863 bursts were detected in 82 hours over 54 days \citep{2022Natur.609..685X}. In the second active episode on September 25-28, 2021, 624 bursts of FRB 20201124A were detected, with a burst being defined as an emission episode where the adjacent emission peaks have a separation shorter than 400 ms \citep{2022RAA....22l4001Z}. These bursts were selected with a frequency resolution of 122.07 kHz, a temporal resolution of $49.152 \,\mu \mathrm{s}$, and a signal-to-noise ratio of $S/N>7$. \cite{2022RAA....22l4001Z} conducted a comprehensive temporal and spectral profile analysis of the bursts detected in the second episode. Some bursts are composed of multiple sub-bursts. These bursts were further divided into several burst events based on their temporal structures, and spectral analysis was performed on these events. A sample of 1268 events was obtained, and their spectra were fitted with a Gaussian function \citep{2022RAA....22l4001Z}. Note that the spectral and temporal analysis of the bursts obtained in the first episode reported by \cite{2022Natur.609..685X} is not done using the same criteria as that for the bursts of the second episode by \cite{2022RAA....22l4001Z}. To avoid the discrepancy between the two samples, we only use the sample of 1268 events reported by \cite{2022RAA....22l4001Z} from the FAST second monitoring campaign for our analysis \footnote{Although the sample selected for analysis was observed in a short temporal coverage in Sep. 2021, it shares similar observational properties with the burst sample observed in June 2021  \citep{2022Natur.609..685X,2022RAA....22l4003J,2022RAA....22l4001Z}}.

We take the duration, the fluence $F_{\nu}^{\rm obs}$, and
and spectral characteristics (including the peak frequency $\nu_p^{\rm obs}$ and the equivalent emission bandwidth $\rm BWe^{\rm obs}$) of the burst events from \cite{2022RAA....22l4001Z}. The data we selected was bandpass calibrated \citep{2022RAA....22l4001Z}. The $\rm BWe^{\rm obs}$ is defined as the full frequency width at 10\% height of the spectral profile that is fitted with a Gaussian function. For the convenience of our following analysis, the $1\sigma$ value ($\sigma_s^{\rm obs}$) and the full-width-half-maximum (FWHM) of the Gaussian spectral function are also given with $\sigma_s^{\rm obs}= \rm BWe/(2\sqrt{2\times \ln{10}})\sim BW_e$/4.29 and $\rm FWHM=2.35\sigma_s^{\rm obs}$. We calculate the burst isotropic energy ($E_{\rm BWe}^{\rm obs}$) over the burst equivalent emission bandwidth ($\rm BWe^{\rm obs}$) in the source rest frame's with
\begin{equation}
E_{\rm BW_{e}}^{\rm obs}
\simeq
\left(10^{39} {\rm erg}\right)
\frac{4 \pi}{1+z}\left(\frac{D_{\rm L}}{10^{28} {\rm ~cm}}\right)^{2}
\left(\frac{F^{\rm obs}_{\nu}}{{\rm Jy} \cdot {\rm ms}}\right)\left(\frac{BW_{e}^{\rm obs}}{\rm GHz}\right).\label{Energy}
\end{equation}

\begin{figure}[!htbp]
\centering
\includegraphics[width=0.5\linewidth]{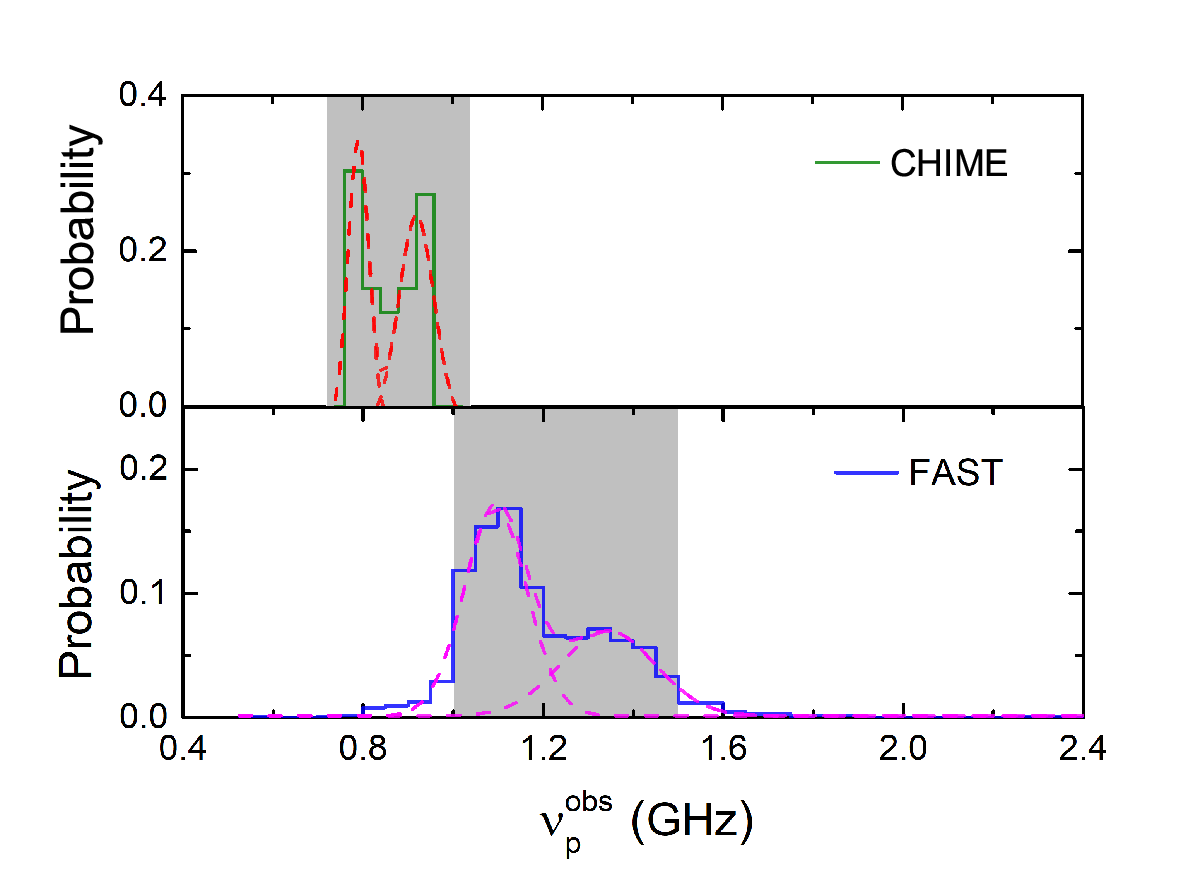}\hspace{-0.1in}
\includegraphics[width=0.5\linewidth]{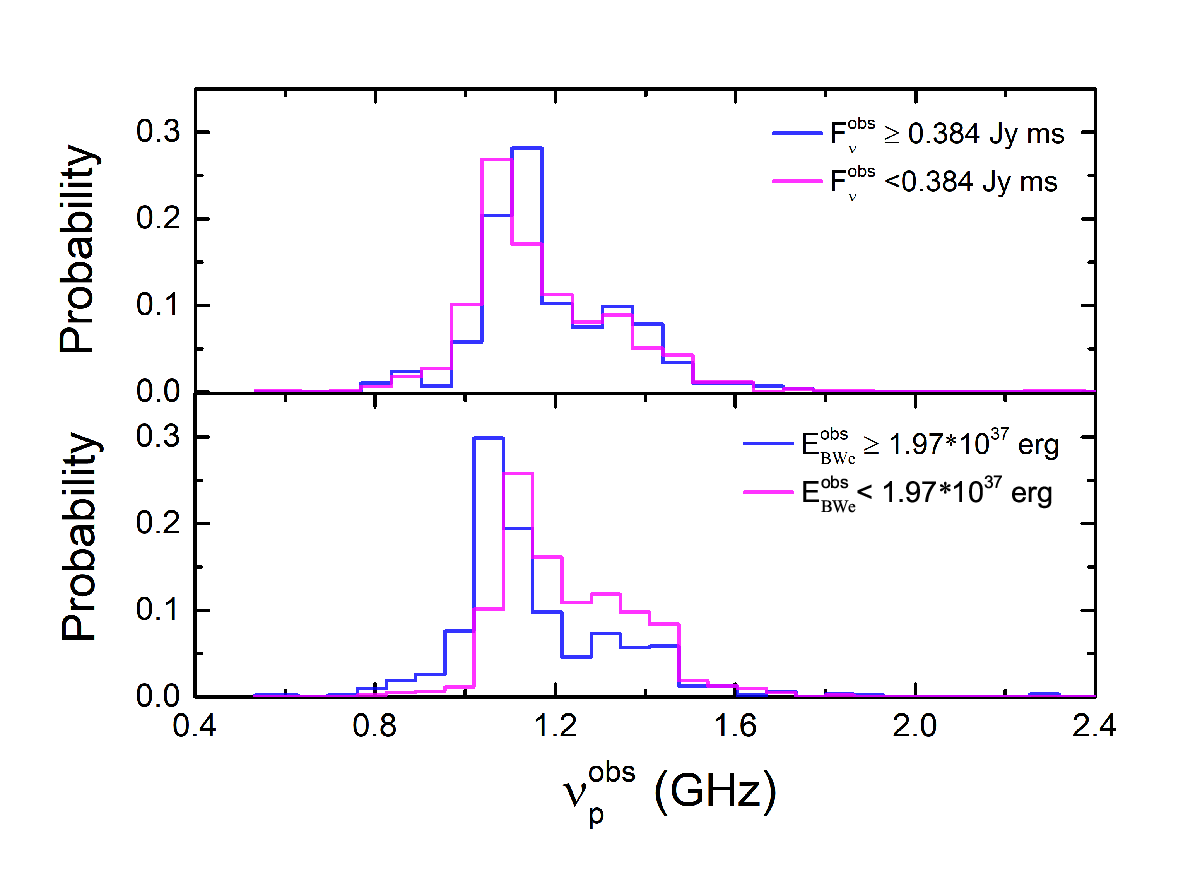}\vspace{-0.1in}
\caption{{\em Left:} The observed peak frequency ($\nu_{p}^{\rm obs}$) distributions of FRB 20201124A in a sample of 1268 burst events (including cluster-bursts) from \cite{2022RAA....22l4001Z} monitored by the FAST telescope in 1.0-1.5 GHz and with a sample of 33 bursts monitored by the CHIME telescope in 0.4-0.8 GHz \citep{2022ApJ...927...59L}. Their profiles are fitted with two Gaussian functions (dashed lines), and the gray shaded areas mark the instrumental bandpasses. {\em Right:} Comparison of the $\nu_{p}^{\rm obs}$ distributions between sub-samples divided by adopting the median ($1.97\times 10^{37}$ erg or $0.384$ Jy ms) of the $E_{\rm BWe}^{\rm obs}$ (or $F_\nu^{\rm obs}$) distribution.
\label{fig:vp_yes_no}}
\vspace{-0.2cm}
\end{figure}

\begin{figure}[!htbp]
\centering
\includegraphics[width=0.45\linewidth]{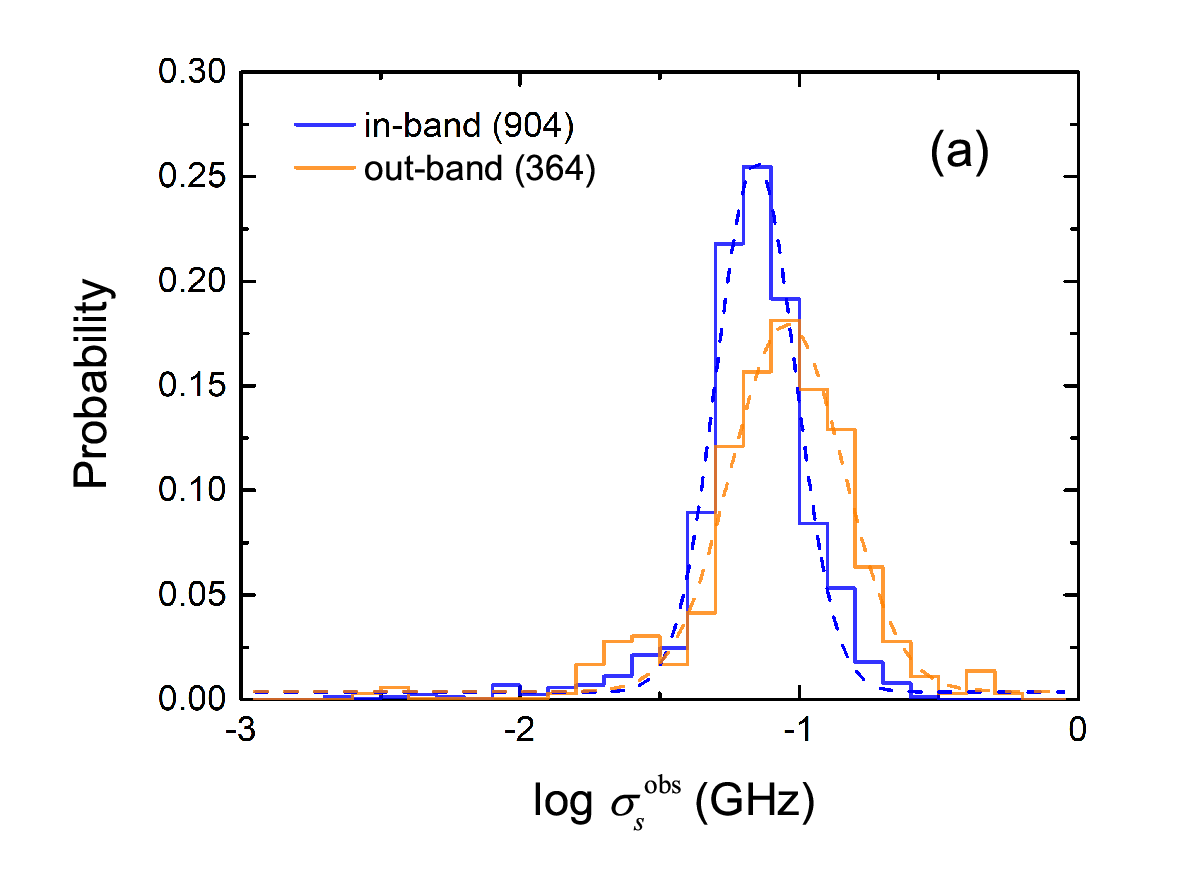}\vspace{-0.1in}
\includegraphics[width=0.45\linewidth]{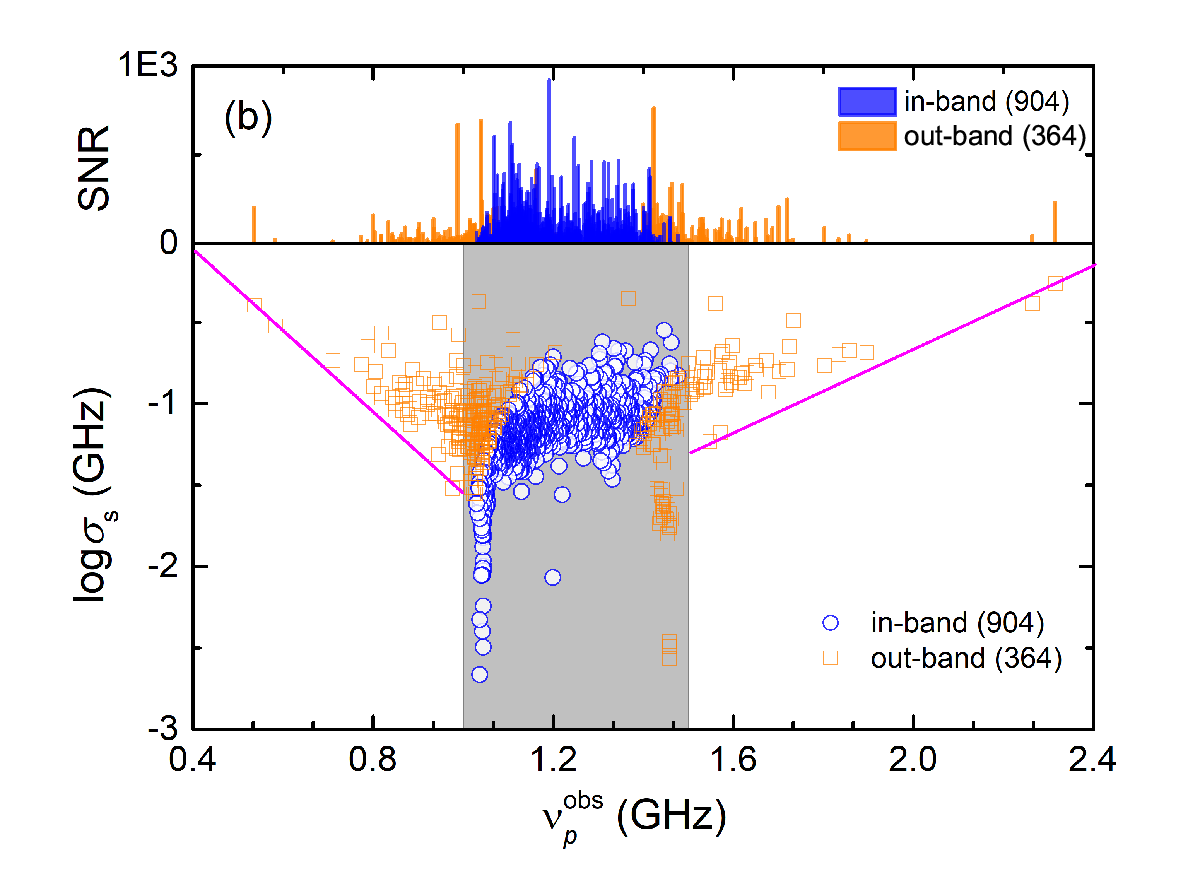}\vspace{-0.1in}
\caption{Comparisons of the one-dimensional $\log \sigma_s^{\rm obs}$ distribution (panel a) and two-dimensional $\log \sigma_s^{\rm obs}-\nu^{\rm obs}_{p}$ distribution (panel b) between the in-band and out-band bursts in our sample for FRB 20201124A observed with the FAST telescope. The dashed lines in panel (a) represent our fits with a Gaussian function. In panel (b), the gray area marks the bandpass of the FAST telescope, and the solid lines mark the detectable limit with the FAST telescope, i.e. a burst cannot be detectable if it is located below the lines.
 \label{fig:sigma}}
\vspace{-0.2cm}
\end{figure}

Figure \ref{fig:vp_yes_no} shows the distribution of $\nu_p^{\rm obs}$ for the bursts in our sample. We find that 90\% (1141 out of 1268) of the burst events have $\nu^{\rm obs}_{p}$ values within the FAST range $[1.0,1.5]$ GHz, indicating that the burst event sample we selected is the burst population that their $\nu^{\rm obs}_p$ values fall within the FAST bandpass. This agrees with that proposed by \cite{2021ApJ...920L..18A} who argued that the peak frequency of most detected bursts should lie in or close to the observational band because otherwise they would not be detected. The bursts with a $\nu^{\rm obs}_{p}$ out of the FAST bandpass should be mostly missed due to this systematic selection effect. Occasionally, a burst that has a $\nu_p^{\rm obs}$ far away from the bandpass could be also detectable if its $\sigma_s^{\rm obs}$ is sufficiently large. We note that two events have $\nu^{\rm obs}_{p}\sim 2.3$ GHz are also detected with $S/N=19$ and $S/N=41$ since their $\sigma_s^{\rm obs}$ are larger than 0.32 GHz (see Figure \ref{fig:sigma}(b)).
The distribution at around $1$ GHz has a sharp drop. The data set and fitting in \cite{2022RAA....22l4001Z} are intrinsically constrained by the observing frequency coverage of the FAST telescope. Examining the burst activity below 1 GHz, it is found that the CHIME telescope detected 33 bursts at 400-800 MHz \citep{2022ApJ...927...59L}. We also show the $\nu^{\rm obs}_{p}$ distribution of these bursts in Figure \ref{fig:vp_yes_no}, where $\nu^{\rm obs}_{p}$ is taken as the middle frequency of the equivalent emission bandwidth. The $\nu_p^{\rm obs}$ distribution of the CHIME sample exhibits two peaks with sharp cut edges. The high-frequency peak is close to the sharp cut at the low-frequency edge of the $\nu_p^{\rm obs}$ distribution for the FAST sample. In addition, the uGMRT telescope detected 48 bursts in 550-750 MHz with a threshold of 1 Jy ms for FRB 20201124A on 2021-April-05 from 12:30 UTC to 16:30 UTC \citep{2022MNRAS.509.2209M}. Combining these observations, we suggest that the sharp cut of the $\nu_p^{\rm obs}$ distribution observed in the FAST sample at the low-frequency end is probably due to the limit of the FAST telescope bandpass.

The $\nu^{\rm obs}_{p}$ of bursts of FRB 20201124A are mostly towards the lower part of the FAST band, but a hump around 1.2-1.5 GHz is also observed in the $\nu_p^{\rm obs}$ distribution. We test the bimodality of the distribution with the KMM algorithm\footnote{The KMM algorithm is a clustering test based probability density estimation approach which can be used to estimate the statistical significance of bimodality in a sample \citep{1994AJ....108.2348A}}. The tests cannot statistically claim the bimodality at a confidence level of $3\sigma$ (chance probability $p<10^{-4}$). Note that the hump is around the center of the FAST bandpass. We suspect whether or not the hump results from the instrument selection for low-energy (or low-flux) bursts having a $\nu_p^{\rm obs}$ close to the center of the bandpass. We divide the whole sample into two subgroups, i.e., low-$E_{\rm BWe}^{\rm obs}$ and high-$E_{\rm BWe}^{\rm obs}$ (or low-$F_\nu^{\rm obs}$ and high-$F_\nu^{\rm obs}$) subgroups, by adopting the median ($1.97\times 10^{37}$ erg or $0.384$ Jy ms) of the $E_{\rm BWe}^{\rm obs}$ (or  $F_\nu^{\rm obs}$) distribution as a division line.
As shown in Figure \ref{fig:vp_yes_no}, an almost identical hump around 1.2-1.5 GHz is observed in the $\nu_p^{\rm obs}$ distributions of both the low- and high-$F_\nu^{\rm obs}$ subgroups. The hump is also observed in the $\nu_p^{\rm obs}$ distributions of the low- and high-$E_{\rm BWe}^{\rm obs}$ subgroups, but the percentages of bursts belonging to the hump are 43$\%$ and 25$\%$ for the two subgroups, respectively. This indicates that low-$E_{\rm BWe}^{\rm obs}$ bursts that have a $\nu_p^{\rm obs}$ value nearer the center of the FAST bandpass have a higher detection rate. Note that the hump does not exactly peak at the central frequency (1.25 GHz) of the FAST bandpass but peaks at $\sim 1.3$ GHz. The slight discrepancy should be due to the statistical fluctuation and/or convolution effects between the intrinsic $\nu_{p}$ distribution and the detection probability. No sharp drop is observed at the upper-frequency end of the FAST bandpass. Note that observations with the 64-m radio dish of Usuda Deep Center$/$JAXA covering the frequency of the S-band (2194-2322 MHz) detected one burst at 2 GHz and no burst was detected at the frequency above 2 GHz with the Deep Space Network \citep{2021ATel14519....1P}. Moreover, all of the five bursts from FRB 20201124A detected with the Parkes UWL receiver, which covers a continuous wide frequency band of 704-4032 MHz, are in the frequency $\leq$ 1.1 GHz \citep{2022MNRAS.512.3400K}. Thus, FRB 20201124A indeed rarely outbursts in the frequency above 2 GHz. Combining these current observations, we infer that FRB 20201124A is active at a frequency range below 2 GHz and mostly toward the lower frequency band of the FAST telescope, making the $\nu_p^{\rm obs}$ distribution peak around 1.0-1.2 GHz.

 Among the 1268 burst events detected with the FAST, 1141 events (90\% bursts) have a $\nu^{\rm obs}_{p}$ value in the FAST range $[1.0,1.5]$ GHz. Furthermore, 906 bursts (71.4\%) satisfy the conditions of $\nu^{\rm obs}_{p}-\sigma_s^{\rm obs}\ge 1.0$ GHz and $\nu^{\rm obs}_{p}+\sigma_s^{\rm obs}\le 1.5$ GHz, indicating that these bursts in the FAST sample cover the major flux ($\ge 68.3\%$). We refer to this subsample as the in-band burst sample and the other subsample of 362 bursts as the out-band burst sample. We compare the distributions of $\log \sigma_s^{\rm obs}$ and show $\log \sigma_s^{\rm obs}$ as a function of $\nu_p^{\rm obs}$ for the in-band and out-band samples in Figure \ref{fig:sigma}. Note that a small fraction (3.8$\%$) of bursts have extremely narrow spectra, i.e. $\sigma_s^{\rm obs}=0.001\sim 0.03$ GHz. The $\nu_p^{\rm obs}$ values of these bursts are close to the edges of the FAST bandpass, as shown in Figure \ref{fig:sigma}(b). This may lead to underestimating their spectral width in the derivation of their $\sigma_s^{\rm obs}$ from the fit with a Gaussian function \citep{2021ApJ...920L..18A,2022RAA....22l4001Z}. In addition, the frequency resolution in the spectral analysis is 122.07 kHz. These bursts may not have enough spectral bins for robustly making spectral analysis. Thus, we exclude these bursts that have $\sigma_s^{\rm obs}<0.03$ GHz in our following analysis.

We fit the $\log \sigma_s^{\rm obs}$ distributions with a Gaussian function and obtain $\log \sigma_{s,c}/{\rm GHz}=-1.16$ ($\sigma_{s,c}=0.07$ GHz, corresponding to an FWHM value of 0.16 GHz) and $\sigma=0.16$ GHz for the in-band burst sample and $\log \sigma_{s,c}/{\rm GHz}=-1.04$ ($\sigma_{s,c}=0.09$ GHz, corresponding to an FWHM value of 0.21 GHz) and $\sigma=0.20$ GHz for the out-band burst sample, where $\log \sigma_{s,c}$ is the central value and $\sigma$ is the standard deviation of the Gaussian function. It is found that the spectra of the in-band bursts are averagely narrower than the out-band bursts. $\sigma_s^{\rm obs}$ is not correlated with $\nu_p^{\rm obs}$. For the out-band bursts, a considerable part of the spectra extends outside the instrumental bandpass, especially for the 10\% bursts whose $\nu^{\rm obs}_{p}$ values are out of the FAST bandpass. Their emission spectra are only partially detected. The detection probability of a burst with $\nu_p$ out of the FAST bandpass depends on the fraction of the observable burst emission that is determined by its $\nu_p$ and $\sigma_s$ values. As shown in Figure \ref{fig:sigma}(b), a few bursts that have a broad spectrum and a $\nu_p^{\rm obs}$ far away from the FAST bandpass can still be detected with a high S/N. By utilizing the bursts that have the largest $\nu^{\rm obs}_{p}$ and smallest $\sigma_s^{\rm obs}$ values among the bursts depicted in Figure \ref{fig:sigma}(b), we demonstrate the selection effect of the FAST telescope observation with violet lines. Bursts below these lines have a $\nu^{\rm obs}_p$ out of the FAST bandpass and a relatively narrow spectral width. They are not detectable with the FAST telescope. This makes a significant observational bias to the out-band bursts in our sample.

\begin{figure}[!htbp]
\centering
\includegraphics[width=0.45\linewidth]{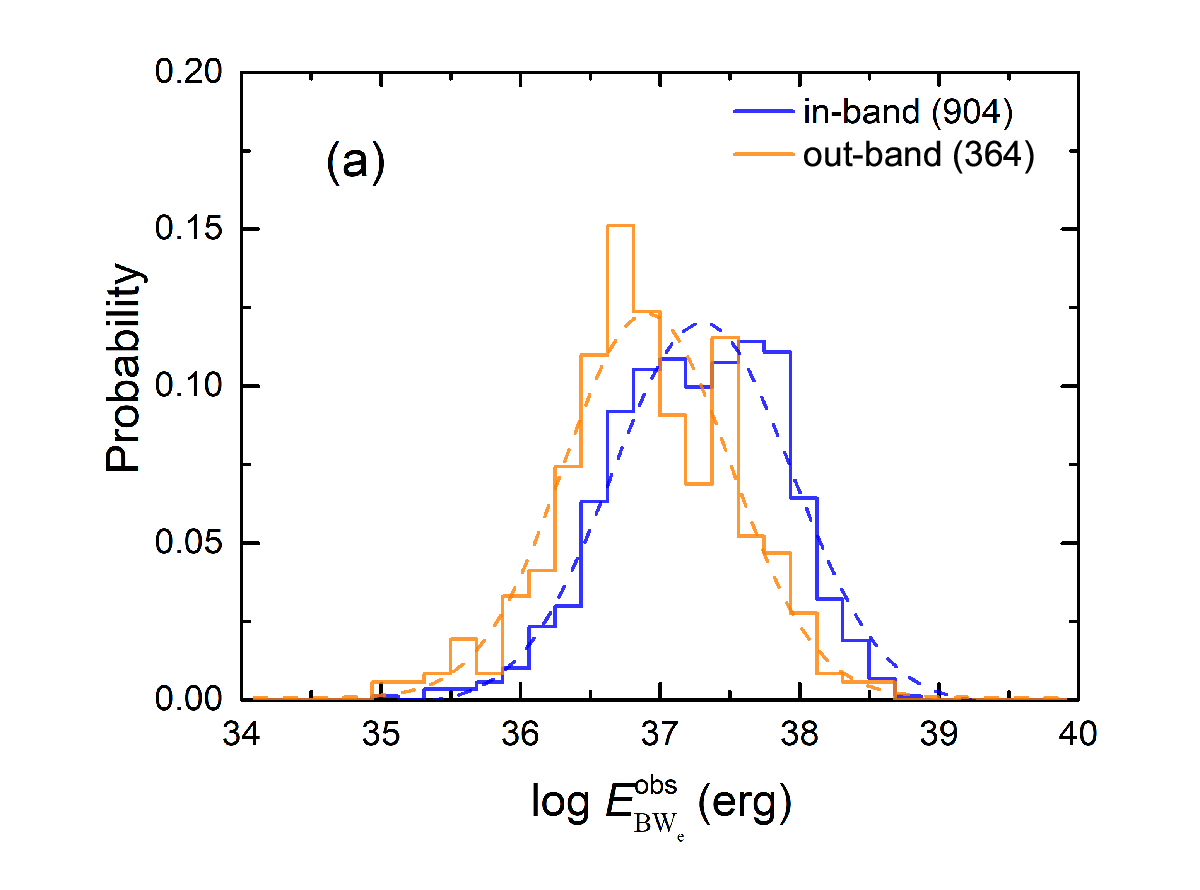}\vspace{-0.1in}
\includegraphics[width=0.45\linewidth]{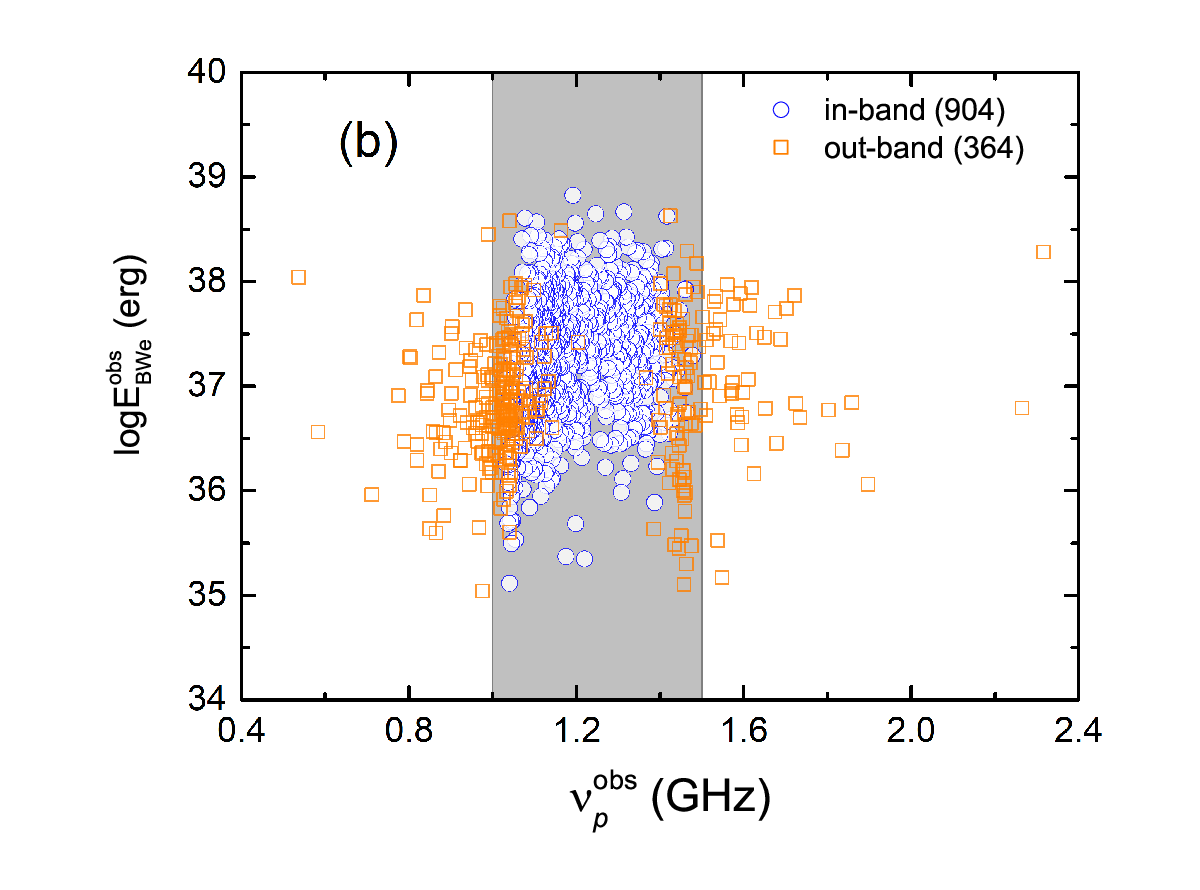}\vspace{-0.1in}
\caption{Comparisons of the one-dimensional $\log E^{\rm obs}_{\rm BWe}$ distribution (panel a) and two-dimensional $\log E^{\rm obs}_{\rm BWe}-\nu^{\rm obs}_{p}$ distribution (panel b) between the in-band and out-band bursts in our sample for FRB 20201124A observed with the FAST telescope. The dashed lines in panel (a) represent our fits with a Gaussian function, and the vertical dash-dotted line is the energy detection threshold, $E_{\rm th}=3.52\times10^{36}\mathrm{erg}$, which is estimated by adopting a fluence threshold of 0.03 Jy ms and the FAST bandwidth (500 MHz) as an equivalent emission bandwidth. The gray area in panel (b) marks the bandpass of the FAST telescope.
\label{fig:E_BWe_obs}}
\vspace{-0.2cm}
\end{figure}

Figure \ref{fig:E_BWe_obs} shows the $\log E_{\rm BWe}^{\rm obs}$ distribution and $\log E_{\rm BWe}^{\rm obs}$ as a function of $\nu_p^{\rm obs}$ for the in-band and out-band samples of bursts. No correlation between $\log E_{\rm BWe}^{\rm obs}$ and $\nu_p^{\rm obs}$ is found. \cite{2022RAA....22l4002Z} proposed that the distribution of isotropic equivalent energy in the burst bandwidth ($ E_{\rm \Delta \nu}^{\rm obs}$) can be fitted with two log-normal functions. We also test the normality of the $\log E_{\rm BWe}^{\rm obs}$ distribution of the whole burst events in \cite{2022RAA....22l4001Z} using the Kolmogorov--Smirnov (K-S) test, Anderson-Darling Normality Test, and D'Agostino's K-squared test. The null hypothesis that they obey a normal distribution is rejected with a probability of $p>10^{-4}$ derived from these tests. Our bimodality test with the Kernel Mixture Modelling (KMM) algorithm cannot statistically claim a bimodality distribution with a p-value of $p_{\rm KS}<10^{-4}$. Comparing the $\log E_{\rm BWe}^{\rm obs}$ distributions between the in-band and out-band samples, one can find that the $E_{\rm BWe}^{\rm obs}$ of the in-band bursts is averagely larger than that of the out-band bursts, i.e. $\log E_{ \rm BWe}^{\rm obs}/{\rm erg}=37.3\pm 0.76$ for the in-band bursts and $\log E_{ \rm BWe}^{\rm obs}/{\rm erg}=36.9\pm 0.72$ for the out-band bursts. The overlap of the $\log E_{\rm BWe}^{\rm obs}$ distributions makes a broad and flat plateau in the range of $\log E_{ \rm BWe}^{\rm obs}/{\rm erg}=36.5\sim 38$. As illustrated in Figure \ref{fig:E_BWe_obs}(b), out-band bursts are detectable only when they are bright and/or broad-spectrum bursts since only a small portion of the spectra fall into the FAST bandpass. Bright bursts with a $\nu_p^{\rm obs}$ far away from the instrument bandpass mostly cannot be detectable. This not only leads to underestimating their burst energies but also creates a significant observational bias on the completeness of the out-band burst sample with the FAST detection threshold. Since the major part of the spectrum of an in-band burst is within the observational bandpass, its $\log E_{ \rm BWe}^{\rm obs}$ value should robustly represent the real burst energy \citep{2017ApJ...850...76L,2021ApJ...922..115A}, and the in-band sample could be regarded as a complete sample considering the FAST telescope detection threshold. Its $\log E_{ \rm BWe}^{\rm obs}$ distribution can be well-fitted with a single Gaussian function, which yields a central value of $\log E_{\rm BWe, c}^{\rm obs}/ \mathrm{erg}=37.3$ and a standard deviation of $\sigma_{\log E_{\rm BWe}^{\rm obs}}=0.76$. The $R^2$ of the fit is 0.97.

\section{Monte Carlo Simulations}
\label{sec:sim}
\begin{figure}[!htbp]
\centering
\includegraphics[width=0.4\linewidth]{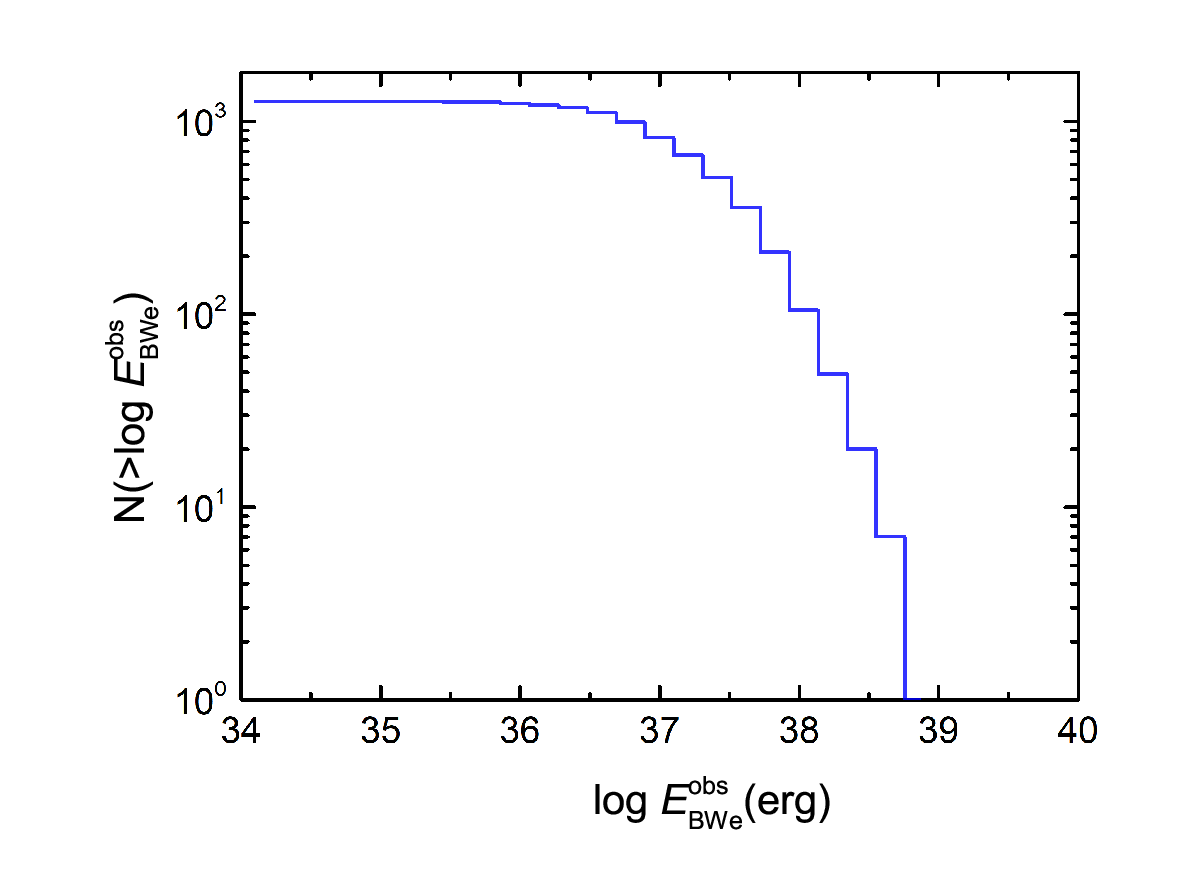}\hspace{-0.1in}
\includegraphics[width=0.4\linewidth]{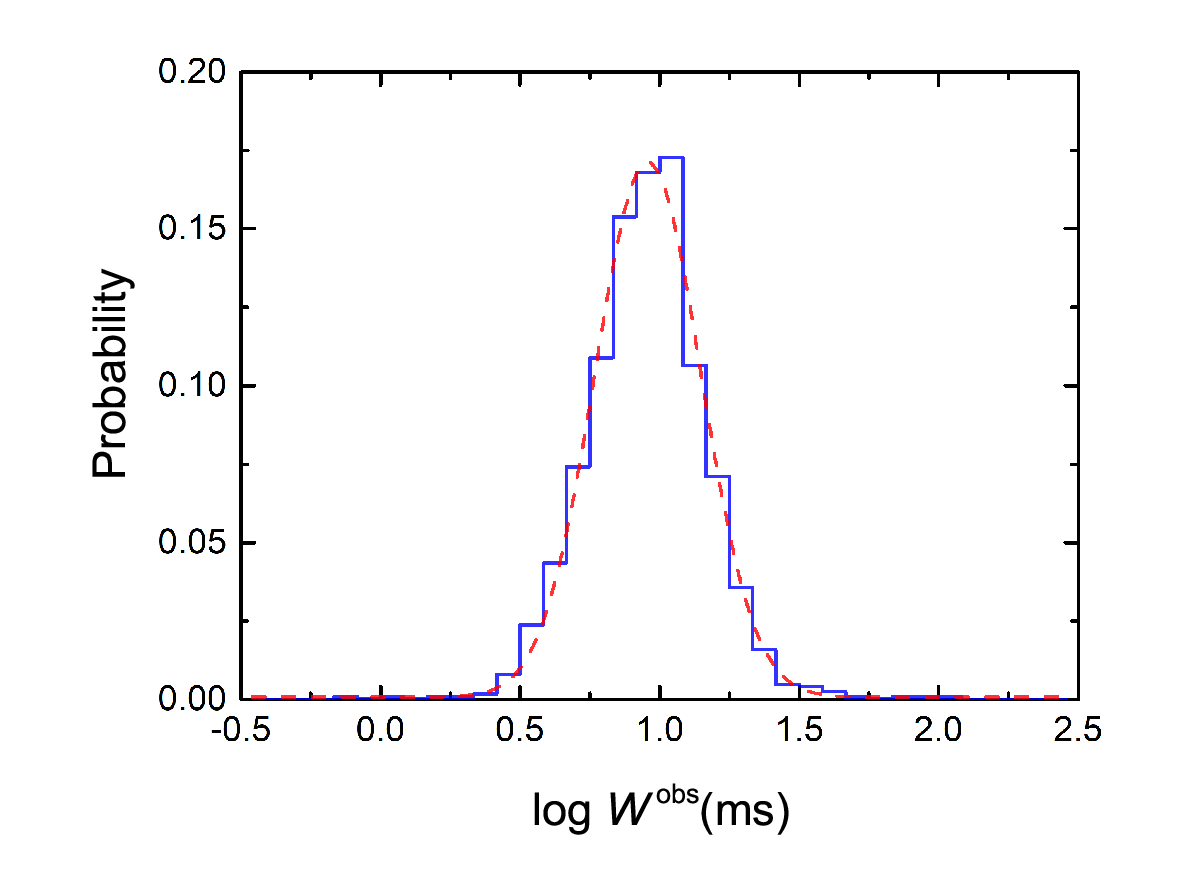}\hspace{-0.1in}
\caption{Observed cumulative distribution of $\log E^{\rm obs}_{\rm BW e}$ (left panel) and the differential distribution of $\log W^{\rm obs}$ (right panel) of our sample of bursts from FRB 20201124A. The dashed line in the right panel represents our Gaussian function fitting of the distribution. \label{fig:duration}}
\vspace{-0.2cm}
\end{figure}

The large and uniform spectral sample of burst events observed with the FAST telescope provides great advantages to constraining the intrinsic spectra and the E-function of FRB 20201124A. Adopting the whole sample of 1268 burst events, we study these issues via Monte Carlo simulations by matching the $E_{\rm BWe}$ distributions and $\nu_p$ distributions of the observed and simulated samples. The cumulative distribution of $\log E_{\rm BW e}^{\rm obs}$ in the sample is displayed in the left panel of Figure \ref{fig:duration}. We describe our empirical model and the procedures of our simulation analysis as follows.

\subsection{Empirical Energy Function and Spectral Models}

First, the intrinsic E-function of FRB 20201124A in the range of $[10^{36},10^{40}]$ erg is adopted as a single power-law (PL) or a cutoff power-law (CPL) function, i.e.
\begin{equation}
\Phi (E) \propto E^{-\alpha^{\rm PL}_E} \label{Eq_p_E_PL}
\end{equation}
or
\begin{equation}
\Phi (E) \propto E^{-\alpha^{\rm CPL}_E} e^{-E/E_c} \label{Eq_p_E}.
\end{equation}
where $\alpha^{\rm CPL}_E$ is the power-law index and $E_c$ is the cut-off energy.

Second, the intrinsic radiation spectrum is modeled with a Gaussian function,
\begin{equation}
F_\nu
=\frac{F}{\sigma_{\rm s} \sqrt{2 \pi}} e^{\frac{-\left(\nu-\nu_{\rm p}\right)^{2}}{2\sigma_{\rm s}^{2}}}, \label{Gauss}
\end{equation}
where $F$ ($=E_{\rm BWe}/4\pi D_L^2$) is the integrated fluence derived from the burst energy $E_{\rm BWe}$, $\nu_{\rm p}$ and $\sigma_{\rm s}$ are the peak frequency and the standard deviation of the intrinsic emission spectrum.
As discussed above, FRB 20201124A is active in the L-band as observed with the FAST telescope. The $\nu^{\rm obs}_{p}$ values of the bursts are mostly toward the lower energy end of the L-band and the $\nu^{\rm obs}_{p}$ distribution seems to peak at $1.0\sim 1.2$ GHz, as exhibited in Figure \ref{fig:vp_yes_no}. Thus, we construct the probability distribution of $\nu_p$ as a Gaussian function,
\begin{equation}
p(\nu_p)
\propto e^{-(\nu_p-\nu_{p,c})^2/2\sigma^2_{\nu_p}},\label{Eq_nu_p}
\end{equation}
where $\nu_{\rm p,c}$ is the center value and $\sigma_{\nu_p}$ is 1$\sigma$ standard deviation. Considering the bursts emitted in the low frequency ($<1$ GHz) \citep{2022ApJ...927...59L,2022MNRAS.509.2209M}, as well as the $\nu_{p}^{\rm obs}$ distribution in the FAST sample, we set $\nu_{\rm p,c}$ uniformly distributed over the range of [0.8, 1.2] GHz and $\sigma_{\nu_p}$$\in [0.05, 0.4]$ GHz. The $\nu_{\rm p,c}$ and $\sigma_{\nu_p}$ values are randomly selected in the corresponding given ranges mentioned above. As presented in Figure \ref{fig:sigma}, the $\sigma^{\rm obs}_s$ values of the in-band bursts are on average smaller than that of the out-band bursts. Both in-band and out-band bursts are detected in a high S/N ratio. Their $\sigma^{\rm obs}_s$ values are robustly estimated by fitting their spectral profiles with a Gaussian function. Therefore, we fit the observed $\sigma^{\rm obs}_s$ distribution of the whole burst events with a log Gaussian function and take it as the intrinsic probability function of $\sigma_s$, i.e.
\begin{equation}
p(\log \sigma_s)
\propto e^{-(\log\sigma_s-\log\sigma_{s,c})^2/2\sigma^2_{\log \sigma_{s}}}, \label{Eq_sigma_s}
\end{equation}
where $\log \sigma_{s,c}/{\rm GHz}=-1.14$, and $\sigma_{\log \sigma_{s}}/{\rm GHz}$=0.17. We should note that $\sigma_{s}$ is not a free parameter in our simulations. It is directly drawn from the distribution of the observed ones based on Eq. (\ref{Eq_sigma_s}).

\subsection{Procedure of Simulations}
As described above, the energy function is set as a PL or a CPL function in our simulation. The spectral profile is described with a Gaussian function with peak frequency $\nu_p$ and standard deviation $\sigma_{\nu_p}$. Both $\nu_{\rm p}$ and $\log \sigma_{\rm s}$ distributions are fixed as a Gaussian function centering at $\nu_{\rm p,c}$ (or $\log \sigma_{\rm s,c}$) with a standard deviation of $\sigma_{\nu_{p}}$ (or $\sigma_{\log \sigma_{s, c}}$). We tabulate prior distributions and ranges of the model parameters in Table \ref{prior_table}. Our simulation procedure is described below.

First, for the PL E-function scenario, $\alpha^{\rm PL}_{E}$ is set to be uniformly distributed in the range of [1, 3]. For the CPL scenario, we set $\alpha_{E}^{\rm CPL}$ and $\log E_{c}$ uniformly distributed in the ranges of $\alpha_{E}^{\rm CPL}\in \{0.3,1.2\}$ and $\log E_{c}/{\rm erg}\in \{37,39\}$, and randomly pick up a set of $\{\alpha_{E}^{\rm CPL}, \log E_{c}\}$.

Second, we generate the $E^{\rm sim}_{\rm BWe}$ value for a given parameter $\alpha^{\rm PL}_{E}$ based on the Eq.(\ref{Eq_p_E_PL}) or for a given set of \{$\alpha_{E}^{\rm CPL},\log E_{c}$\} based on the Eq.(\ref{Eq_p_E}).

Third, we generate a spectral profile for a mock burst event by bootstrapping a set of $\{\nu_{\rm p}$, $\sigma_{s}$\} based on the probability functions of $p(\nu_p)$ and $p(\log \sigma^{\rm obs}_s)$ described by Eqs. (\ref{Eq_nu_p}) and (\ref{Eq_sigma_s}), respectively. The free parameters ($\nu_{\rm p,c}$ and $\sigma_{\nu_{\rm p}}$) of the $p(\nu_p)$ function are randomly picked up from a uniform distribution in the given ranges, i.e. $\nu_{\rm p,c}\in [0.8,1.2]$ GHz and $\sigma_{\nu_{\rm p}}$$\in [0.05,0.4]$. We generate a $\nu_{\rm p}$ value for a given set of $\nu_{\rm p,c}$ and $\sigma_{\nu_{\rm p}}$. Similarly, we generate a $\sigma_{s}$ based on the $p(\log \sigma^{\rm obs}_s)$ function by taking $\log \sigma_{s,c}/{\rm GHz}=-1.14$ and $\sigma_{\log \sigma_{s}}/{\rm GHz}$=0.17, which are derived from our fit to the observed $p(\log \sigma^{\rm obs}_s)$ distribution with a Gaussian function.
\begin{table}
    \centering
\caption{Prior distributions and ranges of the free model parameters.}\label{prior_table}
\begin{tabular}{|l|l|l|l|l|}
\hline
Energy Function & Parameter &Distribution&Range \\
\hline
PL: $\Phi (E) \propto E^{-\alpha^{\rm PL}_E}$ & $\alpha_{E}^{\rm PL}$ &uniform &1.0$\leq\alpha_{E}^{\rm PL}\leq$ 3.0\\
\hline
CPL: & $\alpha_{E}^{\rm CPL}$ & uniform&0.3$\leq\alpha_{E}\leq$ 1.2\\
$\Phi (E) \propto E^{-\alpha^{\rm CPL}_E} e^{-E/E_c}$ & $\log E_{c}^{\rm CPL}/{\rm erg}$ &uniform &37.0 $\leq\log E_{c}\leq$ 39.0\\
\hline
\hline
Gaussian Spectral Function & Parameter &Distribution&Range \\
\hline
$F_\nu=\frac{F}{\sigma_{\rm s} \sqrt{2 \pi}} e^{\frac{-\left(\nu-\nu_{\rm p}\right)^{2}}{2\sigma_{\rm s}^{2}}}$&$\nu_{p,c}/{\rm GHz}$ &uniform&0.80 $\leq\nu_{p,c}\leq$1.20\\
$p(\log \sigma_s)
\propto e^{-(\log\sigma_s-\log\sigma_{s,c})^2/2\sigma^2_{\log \sigma_{s}}}$& $\sigma_{\nu_{p}}/{\rm GHz}$&uniform&0.05$\leq$$\sigma_{\nu_{p}}$$\leq$ 0.40\\
$p(\nu_p)
\propto e^{-(\nu_p-\nu_{p,c})^2/2\sigma^2_{\nu_p}}$& 
$\sigma_{s}/{\rm GHz}$ & a fixed log-normal&-2$\leq\log \sigma_{s}\leq$0\\
 &$\sigma_{s,c}$ &fixed & $\log \sigma_{s,c}/{\rm GHz}=-1.14$\\ &$\sigma_{\log \sigma_{s}}$ & fixed & $\sigma_{\log \sigma_{s}}/{\rm GHz}$=0.17\\
\hline
\end{tabular}
\end{table}

Fourth, we bootstrap a burst duration $W^{\rm sim}$ based on the probability distribution of $p(\log W^{\rm obs})$, which is derived from our Gaussian function fit to the observed burst duration of the whole FAST sample as shown in the right panel of Figure \ref{fig:duration}, i.e.
\begin{equation}
    p(\log  W^{\rm obs})
\propto e^{-(\log  W^{\rm obs}-\log W^{\rm obs}_c)^{2}/2\sigma_{\log  W^{\rm obs}}^{2}},
\label{Eq_w}
\end{equation}
where the best fit estimate yields $\log W^{\rm obs}_c/ \mathrm{ms}=0.96$ and $\sigma_{\log W^{\rm obs}_c}=0.19$ ($R^2=0.99$).

Fifth, we define the observable frequency range $[\lambda_1, \lambda_2]$ for a simulated burst monitored by the FAST telescope with the bandpass [1.0, 1.5] GHz as $\lambda_1=\max(1.0, \nu_1)$ GHz and $\lambda_2=\min(\nu_2, 1.5)$ GHz, where $\nu_1=\nu_{p}-\sigma_{s}$ and $\nu_{2}=\nu_{p}+\sigma_{s}$. The ``observable" specific fluence $F_{\nu}$ is calculated in the range of $[\lambda_1,\lambda_2]$, and the flux density $S_\nu^{\rm sim}$ can be calculated with $S_\nu^{\rm sim}=F_{\nu}/W^{\rm sim}$.

Sixth, we examine whether the burst can be detectable with the FAST telescope. As shown in \cite{2021Natur.598..267L} (Extended Data Figure 5), the completeness of the FAST survey changes with the burst observed fluence and the duration, and the fluence threshold of FRB 20121102A was derived assuming 1 ms burst width in the FAST campaign. We estimate the detection sensitivity of the FAST telescope for FRB 20201124A through the radiometer equation available in \cite{2016MNRAS.458..708C} by adopting the maximum value of the duration in the FAST sample and the following fundamental performance parameters of the FAST from \cite{2020RAA....20...64J,2021ApJ...923..230L}: the system temperature $T_{\rm sys}=25$ K, the telescope gain $G$=16 $\mbox{K Jy}^{-1}$, the telescope bandwidth $B$=500 MHz, the signal-to-noise ratio $<S/N>$=7, and the number of polarizations summed $N_{p}$ =2. This gives the flux density threshold of $S^{\rm FAST}_{\rm th}\sim 1.0\times 10^{-3} \rm Jy$ for FRB 20201124A detected by the FAST. If a simulated burst satisfies the criteria of $S_{\nu}> S^{\rm FAST}_{\rm th}$, it is ``observable". Then, we calculate its $E_{\rm BWe}^{\rm sim}$ value over the equivalent emission frequency range using the parameters of the Gaussian spectral function.

By repeating the second to sixth steps, we generate a simulated sample of 2000 detectable bursts for a given set of \{$\alpha^{\rm PL}_{E}$, $\nu_{\rm p,c}$, $ \sigma_{\nu_{\rm p}}$ \} or \{$\alpha^{\rm CPL}_{E}$, $E_{c}$, $\nu_{\rm p,c}$,$ \sigma_{\nu_{\rm p}}$ \}. Each burst is characterized by a parameter set of $\{E^{\rm sim}_{\rm BWe},\nu^{\rm sim}_p, \sigma^{\rm sim}_s, W^{\rm sim}\}$.

\subsection{Results}
We measure the consistency of the $E_{ \rm BWe}$ and $\nu_{p}$ distributions between the observed and simulated samples using the $p-$value of the K-S test ($p^{E}_{\rm KS}$ and $p^{\nu_p}_{\rm KS}$). The consistency can be claimed in a confidence level of 3$\sigma$ if the $p-$ value is greater than $10^{-4}$. Thus, we consider those parameter sets that yield $p^{E}_{\rm KS}$ and $p^{\nu_p}_{\rm KS}$ values greater than $10^{-4}$ simultaneously to be statistically accepted. It is important to note that the $p^{E}_{\rm KS}$ and $p^{\nu_p}_{\rm KS}$ values are independent. We determine the best model parameter set with the conditional maximum $p$-value as $p^{\max}_{\rm KS}=\max({p^{E}_{\rm KS}}|_{p^{\nu_p}_{\rm KS}>10^{-4}},{p^{\nu_p}_{\rm KS}}|_{p^{E}_{\rm KS}>10^{-4}})$, which yields the maximum $p$-value in a condition of $p^{\nu_p}_{\rm KS}>10^{-4}$ or $p^{E}_{\rm KS}>10^{-4}$.

\begin{figure}[!htbp]
\centering
\includegraphics[width=0.45\linewidth]{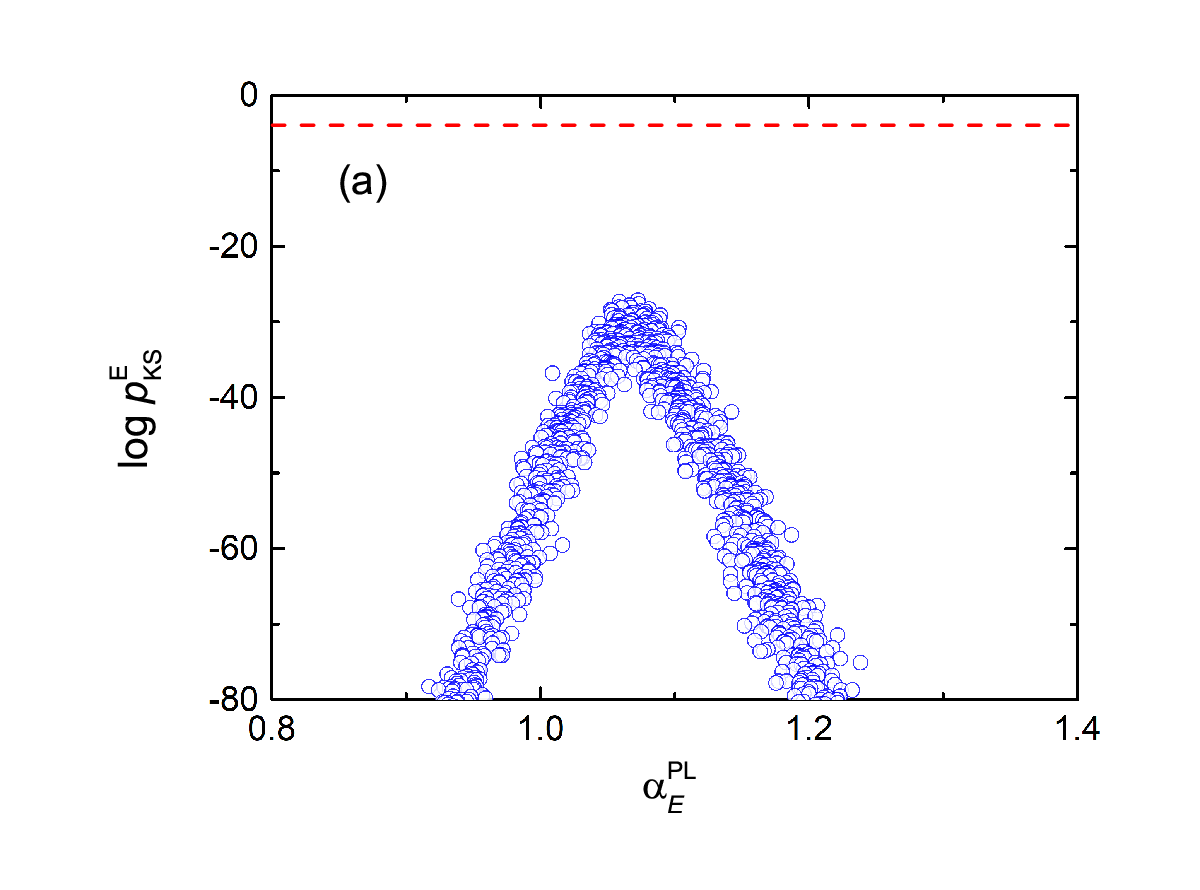}\hspace{-0.3in}
\includegraphics[width=0.45\linewidth]{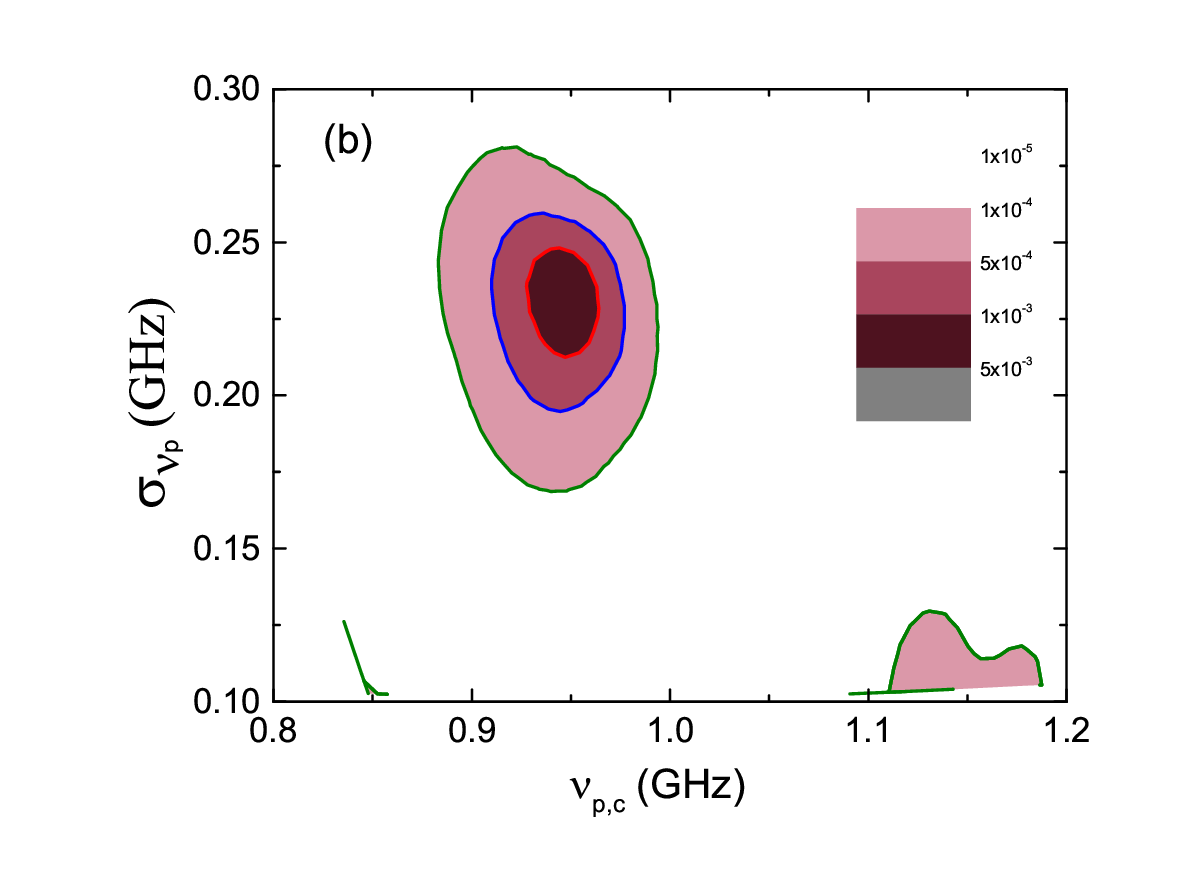}\hspace{-0.3in}
\caption{Panel (a)-$\log p_{\rm KS}^{E}$ as a function of $\alpha_{E}^{\rm PL}$ and panel (b)-contours of $\log p_{\rm KS}^{\nu_p}$ in the $\nu_{p,c}-$$\sigma_{\nu_{p}}$ for FRB 20201124A derived from our simulations by modeling the intrinsic $E$-distribution as a PL function. The red dashed line in panel (a) marks $\log (p_{\mathrm{KS}}^{E})$=-4. }
\label{fig:PL_countour_TO}
\vspace{-0.2cm}
\end{figure}

Figure \ref{fig:PL_countour_TO}(a) displays the $\log p_{\rm KS}^{E}$ as a function of $\alpha_{E}^{\rm PL}$ and Figure \ref{fig:PL_countour_TO}(b) shows the $\log p_{\rm KS}^{\nu_p}$ contours in the $\nu_{p,c}-\sigma_{\nu_{p}}$ plane for the PL energy function scenario. The maximum $p^{E}_{\mathrm{KS}}$ is $=7.60\times 10^{-28}$, and the maximum $p_{\mathrm{KS}}^{\nu_{p}}$ is $5.34 \times 10^{-2}$. We do not find any set of parameters that can yield a conditional maximum $p$-value, indicating that the energy function for the FAST sample of FRB 20201124A in \cite{2022RAA....22l4001Z} cannot be modeled by a PL function.

\begin{figure}[!htbp]
\centering
\includegraphics[width=0.45\linewidth]{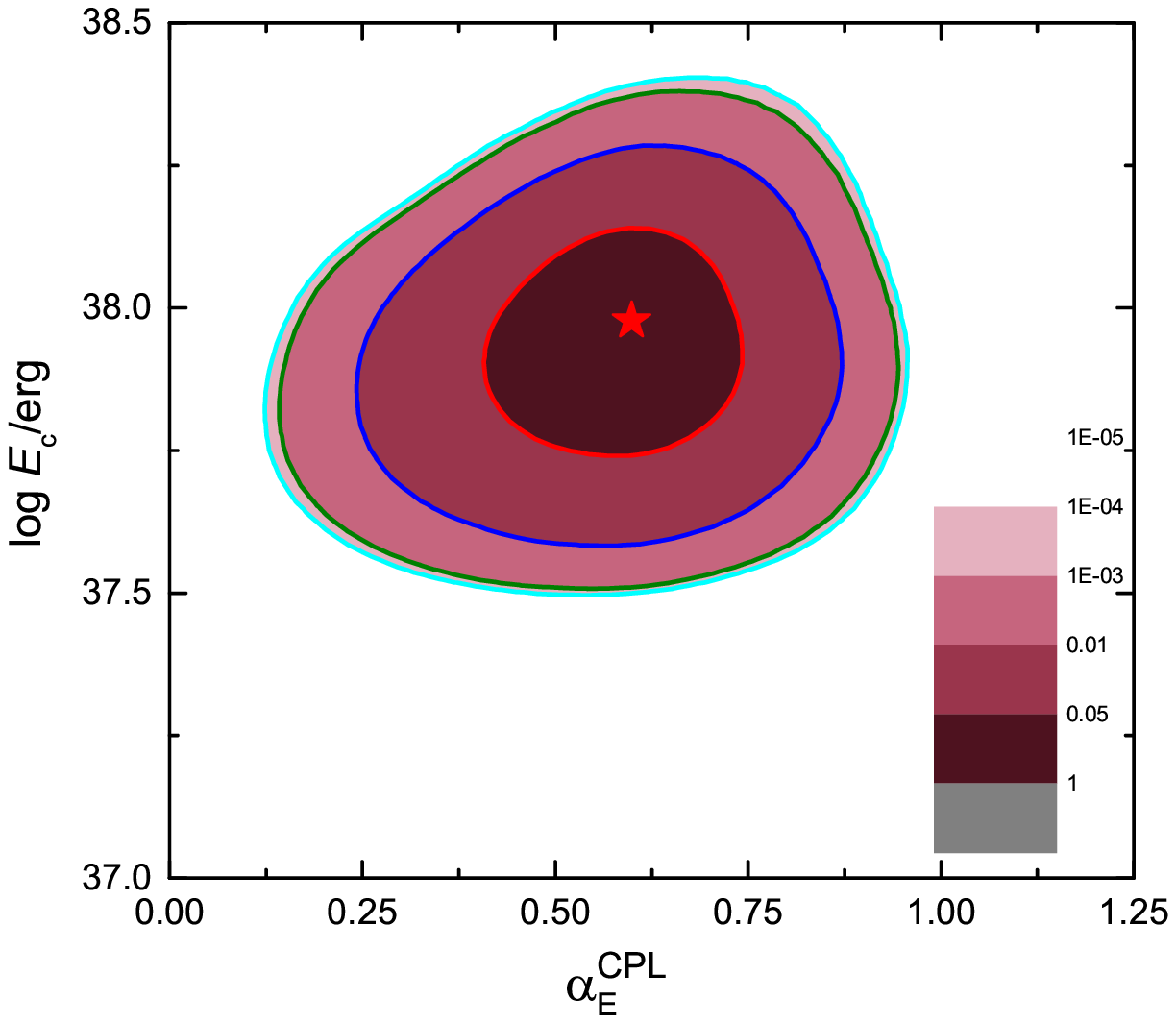}\hspace{-0.3in}
\includegraphics[width=0.45\linewidth]{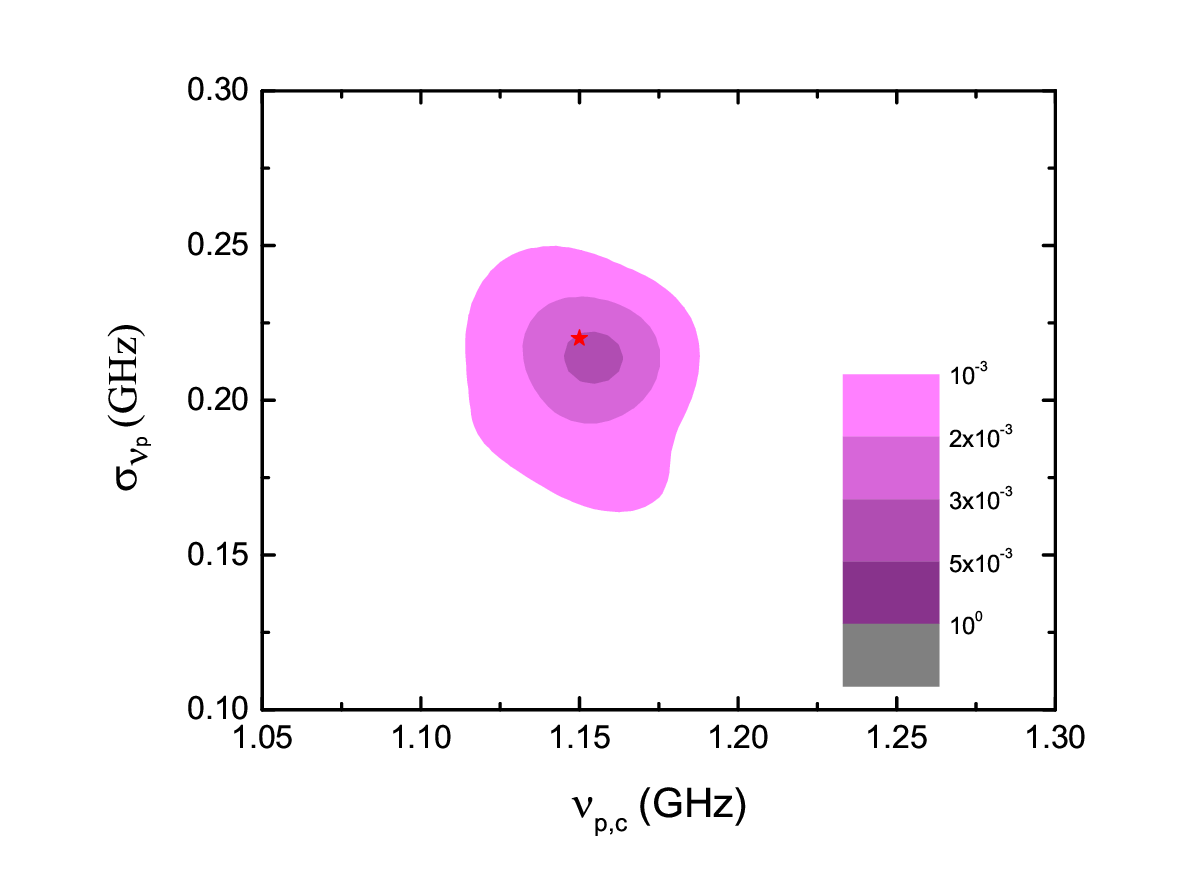}\hspace{-0.3in}
\caption{Contours of $\log p_{\rm KS}^{E}$ in the $\log E_c-\alpha_{E}^{\rm CPL}$ (panel a) and contours of $\log p_{\rm KS}^{\nu_p}$ in the $\nu_{p,c}-\sigma_{p}$ (panel b) for FRB 20201124A derived from our simulations by modeling the $E$-function as a cutoff power-law (CPL) function. The corresponding $p_{\rm KS}$ values of the color legends and boundary lines are marked in each panel. The red star marks the parameter set ($\alpha^{\rm CPL}_E=0.60, \ \log E_c/{\rm erg}=37.98$, $\nu_{\rm p,c}$=1.16 GHz, $\sigma_{\nu_{\rm p}}$=0.22 GHz) that yields that maximum conditional $p^{\rm max}_{\rm KS}$, i.e. $p^{\rm max}_{\mathrm{KS}}=0.65$.
\label{fig:CPL_countour_TO}}
\vspace{-0.2cm}
\end{figure}

\begin{figure}[!htbp]
\centering
\includegraphics[width=0.45\linewidth]{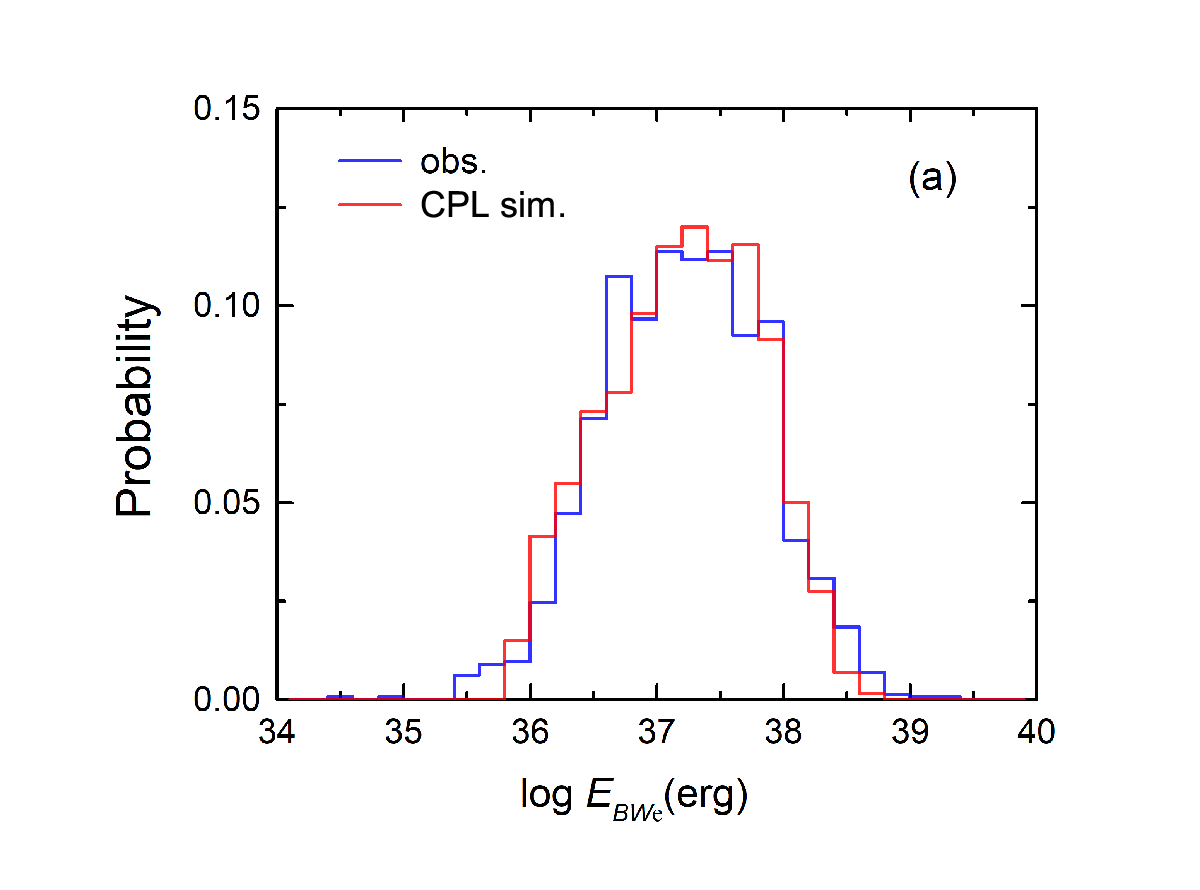}\hspace{-0.3in}
\includegraphics[width=0.45\linewidth]{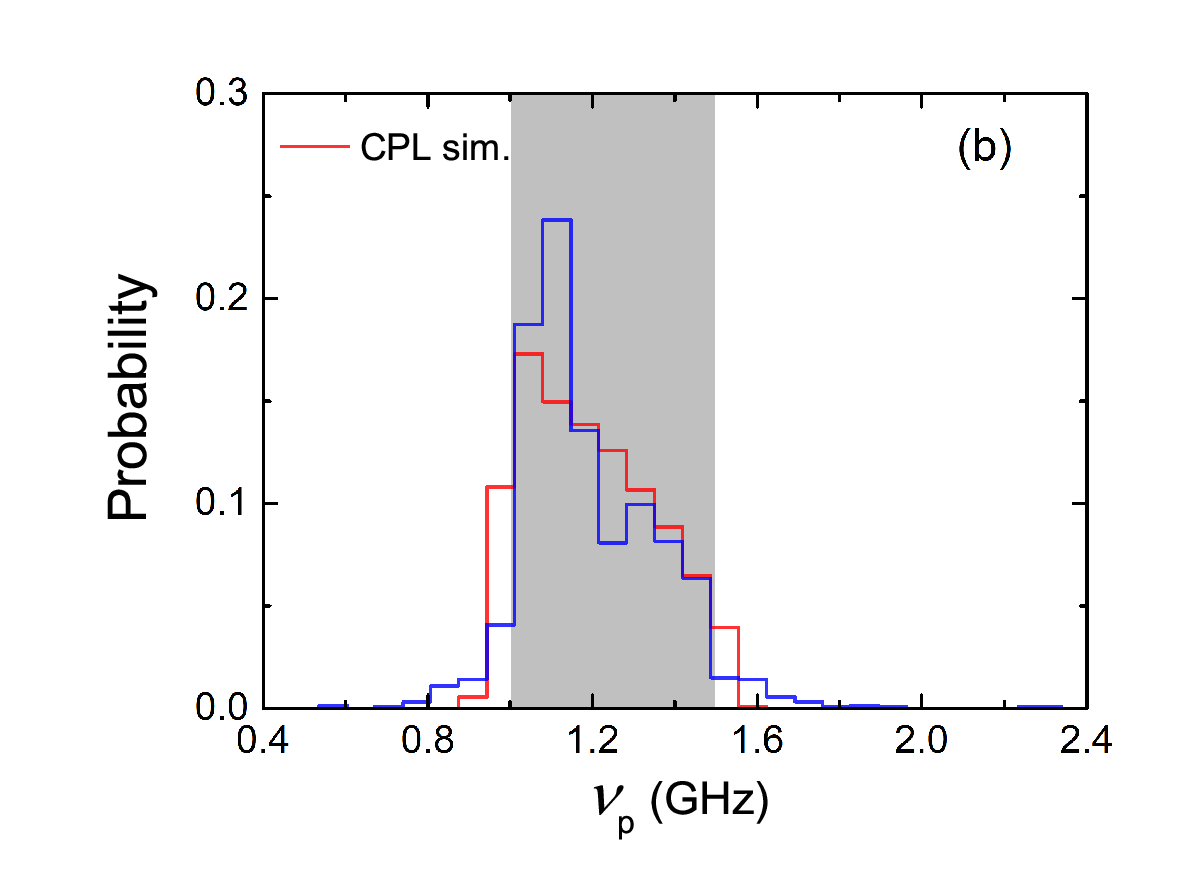}\hspace{-0.3in}
\caption{Comparison of the simulated (red) and observed (blue) distributions of isotropic energy $E_{ \rm BWe}$ (panel a) and the burst peak frequency $\nu_{p}$ (panel b) of FRB 20201124A in the CPL energy function scenario by taking the derived best parameter set ($\alpha^{\rm CPL}_E=0.60, \ \log E_c/{\rm erg}=37.98$, $\nu_{\rm p,c}$=1.16 GHz, $\sigma_{\nu_{\rm p}}$=0.22 GHz) that yields that maximum conditional $p^{\rm max}_{\rm KS}$, i.e. $p^{\rm max}_{\mathrm{KS}}=0.65$, as marked with red stars in Figure \ref{fig:CPL_countour_TO}. The gray area in panel (b) labels the FAST bandpass (1.0-1.5 GHz).
\label{fig:CPL_best_one}}
\vspace{-0.2cm}
\end{figure}

Figure \ref{fig:CPL_countour_TO} illustrates the $\log p_{\rm KS}^{E}$ contours in the $\alpha_E^{\rm CPL}-\log E_c$ plane and the $\log p_{\rm KS}^{\nu_p}$ contours in the $\nu_{p,c}-\sigma_{\nu_{p}}$ plane for the CPL energy function scenario. With the criteria of $p_{\rm KS}^{\rm E}>10^{-4}$, we have $\log E_c/{\rm erg} \in [37.50, 38.40]$ and $\alpha_E^{\rm CPL}\in [0.12, 0.96]$.
The color contours of Figure \ref{fig:CPL_countour_TO}(b) show the constraints on the spectral parameters. We find that $ \nu_{p,c}<1.2$ GHz can yield $p^{\nu_{p}}_{\rm KS}>10^{-4}$, as marked with in Figure \ref{fig:CPL_countour_TO}(b). The derived  $p^{\max}_{\rm KS}$ is 0.65 and the corresponding best parameter set is $\{\alpha_E^{\rm CPL}=0.60,\log E_c/{\rm erg}=38.0, \nu_{\rm p,c}=1.16 {\rm ~ GHz}, $$\sigma_{\nu_{\rm p}}$$=0.22 {\rm ~GHz}\}$.
Figure \ref{fig:CPL_best_one} displays the comparison between the observed and simulated $E^{\rm sim}_{\rm BWe}$, $\nu^{\rm sim}_{p}$, and $\sigma^{\rm sim}_s$ distributions by adopting this parameter set. It is found that the $E^{\rm obs}_{\rm BWe}$ distribution can be well reproduced, and the observed $\nu_{p}^{\rm obs}$ distribution is also globally reproduced. It is worth noting that the hump at around $1.2\sim 1.3$ GHz in the $\nu_p^{\rm obs}$ distribution is not reproduced in our simulations. This might also imply that some systematic effect, such as the high detection probability of events with $\nu_p^{\rm obs}=1.2\sim 1.3$ GHz, is not modeled by our simulations.

\section{Discussion}
\label{sec:Discuss}

FRB 20201124A shares similar observational properties with the typical repeating FRB sources- FRB 20121102A and FRB 20190520B. For example, some of their bursts exhibit sub-burst drifting, narrow-band limited emission, high linear polarization, and large circular polarization \citep{2021Natur.598..267L,2022MNRAS.515.3577H,2022RAA....22l4001Z,2023MNRAS.522.5600L}.  
We compare its peak frequencies ($\nu^{\rm obs}_p$ and inferred $\nu_p$) distributions in the observer's frame and the energy function to the other two FRBs and discuss possible implications for radiation physics.

\subsection{\texorpdfstring{$\nu_{p}$}. Distribution}
\begin{figure}[!htbp]
\centering
\includegraphics[width=0.65\linewidth]{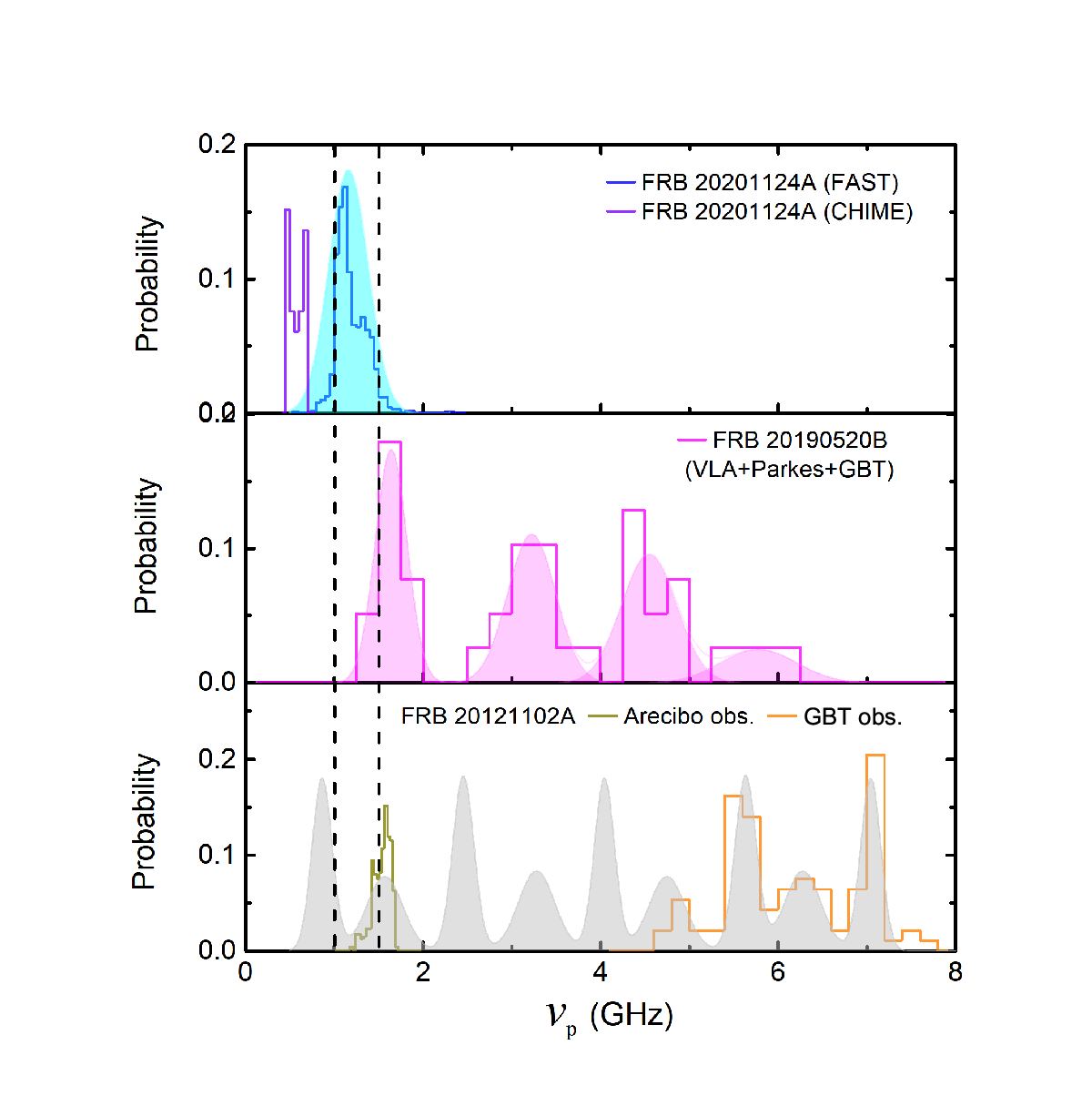}\vspace{-0.1in}
\caption{Comparison of the observed (lines) and inferred intrinsic (shaded areas) $\nu_{p}$ distributions of FRB 20201124A with FRB 20121102A \citep{2022ApJ...941..127L} and FRB 20190520B \citep{2023MNRAS.522.5600L}. The dashed lines mark the bandpass of the FAST.
\label{fig:vp_333}}
\vspace{-0.2cm}
\end{figure}

Our above analysis show that the burst $\nu^{\rm obs}_{\rm p}$ distribution of FRB 20201124A is in the range of 0.4-2 GHz and peaks around 1 GHz. The observed spectral properties of the repeating FRBs are heavily influenced by the various observational biases, such as the sensitivity limit of the instrument, and the observing band \citep{2017ApJ...850...76L,2021ApJ...920L..18A}. We compare $\nu^{\rm obs}_{\rm p}$ of FRB 20201124A with that of FRB 20121102A and FRB 20190520B presented by \cite{2022ApJ...941..127L,2023MNRAS.522.5600L}, who reported the observed $\nu_{p}^{\rm obs}$ distributions of these two FRBs by utilizing the data gathered from multiple telescopic observation campaigns.
As shown in Figure \ref{fig:vp_333}, the $\nu^{\rm obs}_{p}$ values of FRB 20121102A and FRB 20190520B are across a broad frequency coverage, i.e. 0.5-8 GHz for FRB 20121102A and $\sim$1-6 GHz for FRB 20190520B, and their distributions shows several distinct peaks \citep{2022ApJ...941..127L,2023MNRAS.522.5600L}. This is dramatically different from that of FRB 20201124A.

The inferred $\nu_p$ distribution of FRB 20201124A is a normal distribution centering at $\nu_{\rm p,c}=1.16{\rm ~ GHz}$ with a standard deviation of 0.22 GHz. Thus, the probability of a $\nu_p$ value falling in the bandpass of the FAST telescope (1-1.5 GHz) is 70.6\%. In the high-frequency end above the FAST telescope bandpass, the probability of detection of a burst with $\nu_p^{\rm obs}>2$ GHz is smaller than $7\times 10^{-5}$. In the low-frequency end below the FAST telescope bandpass, the probability of a $\nu_p$ in the range of 0.4-1 GHz is 0.23, suggesting a considerable fraction of bursts are detectable with the CHIME and Parkes telescopes as reported by  \cite{2022ApJ...927...59L} and \cite{2022MNRAS.512.3400K}.

The $\nu_p$ distribution of FRB 20190520B has several peaks across the frequency range of 1-6 GHz. Thus, it was detected with the FAST, VLA, GBT, and Parkes telescopes. It is not very active in the FAST bandpass in comparison with FRB 20201124A, i.e. $\sim$4.5 $\rm h^{-1}$ for FRB 20190520B \citep{2022Natur.606..873N} vs. 542 $\rm hr ^{-1}$ \citep{2022RAA....22l4001Z}. One $\nu_p$ distribution peak $\nu_{\rm p,c}$ is at $1.64\pm0.18$ GHz \citep{2023MNRAS.522.5600L}, which is the nearest one to the bandpass of the FAST telescope. However, the probability of $\nu_p$ within the bandpass is only 0.22. Due to the narrow-banded emission of FRB 20190520B, only the bursts of this peak can be detectable with the FAST telescope. The event rate of FRB 20190520B in 1-2 GHz inferred from the FAST observations might therefore be underestimated. In addition, most of the bursts detected with the FAST telescope would be out-band bursts since the $\nu_{\rm p,c}$ is beyond the bandpass. This is confirmed by the fact that the $\nu^{\rm obs}_p$ values of the bursts from FRB 20190520B observed with the FAST telescope cannot be well constrained \citep{2022Natur.606..873N}.

The $\nu_p$ distribution of FRB 20121102A spans across the frequency range of 0.5-8 GHz. It was extensively detected with the FAST, VLA, GBT, Arecibo, Parkes telescopes, etc. It is very active in the FAST bandpass. The FAST bandpass is situated between the two adjacent peaks ($0.87\pm 0.12$ GHz and $1.57\pm 0.21$ GHz) of the $\nu_p$ distribution \citep{2022ApJ...941..127L}. The accumulated probability of the first $\nu_p$ distribution peak within the FAST bandpass is $0.14$, and it is 0.37 for the second $\nu_p$ distribution peak. Their $\nu_{p,c}$ values are very close to the low/high-frequency edge of the FAST bandpass, with deviations of $0.13$ and $0.07$ GHz. Thus a large fraction of bursts with $\nu_p$ being out of the bandpass can be detectable with the benefit of the high sensitivity of the FAST telescope. This should be why the spectral peak frequencies of the bursts observed with the FAST telescope reported by \cite{2021Natur.598..267L} are not available. The $\nu_p$ distribution peak at $1.57\pm 0.21$ GHz well covers ($\sim 75\%$) the bandpass of the Arecibo telescope, leading to the $\nu_p$ of bursts from FRB 20121102A observed with the  Arecibo telescope to be robustly estimated \citep{2022MNRAS.515.3577H}.

\subsection{Observed Energy Distribution and Intrinsic Energy Function}
As discussed above, the observed burst energy and the intrinsic energy function of the bursts from the three FRBs depend on the instrumental bandpass due to their narrow emission spectrum and the fringe $\nu_p$ distributions  \citep{2021ApJ...920L..18A,2021ApJS..257...59C,2022ApJ...941..127L,2023ApJ...947...83A,2023MNRAS.522.5600L}. Taking FRB 20121102A as an example, the isotropic energy distribution derived from observations with the FAST telescope in the range of 1-1.5 GHz is a tentative bimodal distribution \citep{2021Natur.598..267L}, but it is clustered at $10^{38}\sim 10^{39}$ erg with the observations of the Arecibo telescope in the range of 1.15-1.73 GHz \citep{2022MNRAS.515.3577H}. The inferred E-function of FRB 20121102A from the observations with the FAST telescope is a single power-law function with an index of $-1.82^{+0.10}_{-0.30}$ \citep{2022ApJ...941..127L}.

The large and uniform $\nu_p$-in-band burst sample of FRB 20201124A should be perfect for studying its E-function. The derived E-function from our analysis is a CPL function with an index of $\alpha_E^{\rm CPL}=0.60$ and a cutoff energy of $E_{c}=9.49 \times 10^{37}$ erg. We compare it with that of FRB 20121102A and FRB 20190520B derived from the FAST observations in Table \ref{table}. One can find that the E-function of FRB 20201124A resembles that of FRB 20190520B, which is also described with a CPL function with $\alpha_E^{\rm CPL}=0.47$ and $E_{c}=7.4 \times 10^{37}$ erg \citep{2023MNRAS.522.5600L}. Different from that of FRB 20201124A and 20190520B, the E-function of FRB 20121102A is modeled with a single power-law with an index of $1.82^{+0.10}_{-0.30}$ \citep{2022ApJ...941..127L}. The observed bimodal energy distribution in \cite{2021Natur.598..267L} and the single PL energy function in \cite{2022ApJ...941..127L} of FRB 20121102A derived from the observation with the FAST telescope may be due to overestimating the burst number at the low energy end with the detection of a large fraction of out-band bursts since the bandpass of the FAST telescope spans across only the valley of two adjacent $\nu_p$ distribution peaks in the frequency range of 0.5-2 GHz.

\subsection{Implication for Radiation Physics}
 The facts that FRB 20201124A is active in a narrow frequency range of 0.4-2.0 GHz and its radiation spectrum normally peaks at $\nu_{p, c}\sim1.16$ GHz likely give insights into the radiation physics of FRBs. These facts, together with the observed diverse polarization swing, the high circular polarization, and the down-drifting disfavor the synchrotron maser emission at highly magnetized shocks for FRB 20201124A \citep{2014MNRAS.442L...9L,2019MNRAS.485.4091M}. In the weakly magnetized plasma, negative absorption can occur below the generalized Razin frequency and generate an extremely narrow spectrum \citep{2017ApJ...842...34W}. Our analysis implies that the plasma conditions of this FRB should be uniform and fine-tuned for generating bursts at $\nu_{\rm p, c}$. Both $\nu_{\rm p, c}$ and $E_c$ should place a tight constraint on the Razin frequency of maser emission in the weakly magnetized plasma scenario \citep{2017ApJ...842...34W}. FRB 20121102A and FRB 20190520B exhibit outbursts in some specific frequencies across a broad frequency range \citep{2022ApJ...941..127L,2023MNRAS.522.5600L}. This likely suggests that these bursts from these three FRBs are generated in a clumpy plasma, which can offer the conditions for maser emission at several specific frequencies. As proposed by \cite{2018MNRAS.477.2470L}, a plasma in the twisted magnetosphere of a magnetar may be clumpy due to two-stream instability.

\begin{table}
    \centering
\caption{Comparison of the derived parameter(s) of the E-function of FRB 20201124A, FRB 20121102A, and FRB 20190520B from the FAST observations through our Monte Carlo simulations.}\label{table}
\begin{tabular}{|c|c|c|c|}
\hline
Source & $\alpha_{E}$ ($\alpha_E^{\mathrm{CPL}}$) & $E_{c}$$\left(10^{37} \mathrm{erg}\right)$&Reference\\
\hline
FRB 20121102A &$1.82_{-0.30}^{+0.10}$ & -&\cite{2022ApJ...941..127L}\\
\hline
FRB 20190520B &  0.47 & 7.40&\cite{2023MNRAS.522.5600L}\\
\hline
FRB 20201124A& 0.60 &9.49& This work \\
\hline
\end{tabular}\\
\end{table}

\section{Summary}
\label{sec:Summary}
By analyzing a large and uniform sample observed with the FAST telescope and some bursts detected by the CHIME telescope, we investigate the spectral property and energy function of FRB 20201124A. We summarize our results as follows.
\begin{itemize}
    \item Combining observations across a broad frequency range (144 MHz-8502 MHz), we show that FRB 20201124A seems to be active in the frequency range of 0.4-2 GHz, and non-detections for this FRB were observed at frequencies outside this frequency range. The observed $\nu_p^{\rm obs}$ values of the emission spectra are mostly toward the lower end of the bandpass of the FAST telescope, illustrating a sharp peak at $\sim 1.1$ GHz and a broad hump at 1.2-1.4 GHz in the $\nu^{\rm obs}_{p}$ distribution observed with the FAST telescope. The hump is likely due to a relatively high detection rate for the bursts that have a low burst energy.
    \item Among the whole sample of all 1268 burst events obtained in \cite{2022RAA....22l4001Z} for our analysis, 904 bursts (71.4\%) are in-band bursts whose $1\sigma$ spectral regimes are within the FAST bandpass (1-1.5 GHz). The spectra of the in-band bursts are averagely narrower than the out-band bursts, i.e. $\log \sigma_{s,c}/{\rm GHz}=-1.16\pm 0.16$ of the in-band bursts and $\log \sigma_{s,c}/{\rm GHz}=-1.04\pm 0.20$ of the out-band bursts. The observed energy of an in-band burst should be a robust representative of the intrinsic burst energy. We obtain $2.0\times10^{37}\mathrm{erg}$ in average for the in-band bursts.
    \item We constrain the intrinsic $\nu_p$ distribution and the energy function of FRB 20201124A via Monte Carlo simulations by matching the observed and simulated $E_{\rm BWe}$ distributions with the whole sample of 1268 bursts. The derived energy function is $\Phi(E)\propto E^{-0.60}e^{-E/E_c}$, where $E_c=9.49 \times 10^{37}$ erg. The derived intrinsic $\nu_p$ distribution is described as a normal function with a center value of $\nu_{p,c}=1.16$ GHz and a standard deviation of $0.22$ GHz.
    \item We compare our results with that of typical repeating FRBs: FRB 20121102A and FRB 20190520B that are active over a broad frequency range at several specific frequencies and discuss possible observational biases on the estimation of the event rate and energy function.
    \item The observed narrow emission spectrum, together with the very high total degree of polarization, suggests that the observed spectra with the FAST telescope should represent the intrinsic one for FRB 20201124A. The maser emission in a weakly magnetized plasma has the potential to explain the narrow spectrum. The narrow-banded bursts with an extremely narrow spectrum likely imply that the conditions of the plasma are uniform and fine-tuned for generating bursts at a specific frequency $\nu_{\rm p, c}=1.16$ GHz with burst energy cutting off at $E_c\sim 10^{38}$ erg. This is dramatically different from the bursts of FRB 20121102A and FRB 20190520B. The two FRBs' narrow spectrum and separated $\nu_p$ distribution peaks suggest that their plasma can offer fine-tuned conditions for maser emission at several specific frequencies.
\end{itemize}

\section*{acknowledgments}
We very much appreciate the very valuable comments and suggestions of the referee. We also thank the helpful discussions with Heng Xu, Yao Chen, Pei Wang, Wei-Yang Wang, and Yuan-Pei Yang. We acknowledge the use of the public data from the FAST/FRB Key Project. This work is supported by the National Natural Science Foundation of China (grant Nos.~11988101 and 12133003). F.L. acknowledges the support from the National SKA Program of China (grant No. 2022SKA0130103), and E. W. L. is also supported by the Guangxi Talent Program (``Highland of Innovation Talents"). D.L. is a  New. Cornerstone Investigator.

\end{document}